\documentclass[%
reprint,
amsmath, amssymb,
aps,
prb,
]{revtex4-2}
\usepackage{graphicx}
\usepackage{dcolumn}
\usepackage{palatino}
\usepackage{bm}
\usepackage{amsmath} 
\usepackage{amssymb} 
\usepackage[utf8]{inputenc} 
\usepackage{textgreek}      
\usepackage{natbib}
\usepackage{hyperref}
\hypersetup{
	colorlinks = true,
	linkcolor = blue,
	citecolor = blue,
	urlcolor = blue
}
\usepackage{natbib}
\usepackage{chemformula}

\begin{document}
	\hyphenpenalty=10000
	\exhyphenpenalty=10000
	\tolerance=1000
	\emergencystretch=3em
	\preprint{APS/123-QED}
	
	\title{First-principles Investigation of CaV$_2$TeO$_8$: A Multifunctional Heteroanionic Oxychalcogenide for Photocatalytic and Thermoelectric Applications}
	\author{G. Thamizharasan}
	\affiliation{Department of Physics, School of Advanced Sciences, Vellore Institute of Technology, Chennai - 600127, India}
	
	\author{R. D. Eithiraj}
	\thanks{Corresponding author: R.D. Eithiraj\\ \href{mailto:your.email@example.com}{eithiraj.rd@vit.ac.in}}
	\affiliation{Department of Physics, School of Advanced Sciences, Vellore Institute of Technology, Chennai - 600127, India}
\date{\today}
	\begin{abstract}
		The development of industrial growth raises global energy demand. Novel materials with promising unorthodox features satiate and demand for sustainable, non-toxic, and cost-effective energy. A first-principles investigation were carried out to ground state and transport properties of \(\mathrm{CaV_2TeO_8}\) towards energy harvesting and photocatalytic applications was carried out. In groundstate the \(\mathrm{CaV_2TeO_8}\) showed good mechanically stable with a volume of \(1138.235 \, \text{\AA}^3\) and a high elastic modulus of \(215.73 \, \text{GPa}\). Further, the electronic properties of \(\mathrm{CaV_2TeO_8}\), PBE-GGA exchange-correlation was utilized, demonstrates direct and indirect transitions with a wide bandgap of 2.7 and 2.8 eV, indicating its semiconducting nature. Photoinduced charge carriers were investigated through Deformation Potential Theory (DPT) to estimate the mobility of charge carriers, excitonic radius, and effective mass. Axial anisotropic strain in band degeneracy and shifts in the S-Y and Y-$\Gamma$ transitions offer a \(3.17 \, \text{\AA}\) Frenkel-type strongly bound exciton with a high carrier mobility of \(2409.91 \, \text{cm}^2 \, \text{V}^{-1} \, \text{s}^{-1}\) electrons and \(316.46 \, \text{cm}^2 \, \text{V}^{-1} \, \text{s}^{-1}\) holes respectively. To estimate the feasibility of \(\mathrm{CaV_2TeO_8}\) in photocatalysis, we examined the conduction and valence band edges under harsh pH conditions within PBE-GGA exchange-correlation functionals. The band edges of \(\mathrm{CaV_2TeO_8}\) show the halfway reaction for hydrogen evolution with reference to the Normal Hydrogen Electrode (NHE). Moreover, the Boltzmann equation interface was used to investigate the transport and thermoelectric properties. A high Seebeck coefficient in \(\mathrm{CaV_2TeO_8}\) at \(300 \, \text{K}\) directs a thermoelectric figure of merit of \(0.94\).
	\end{abstract}
    \maketitle
\section{\label{sec:level1}Introduction}
	The development of humanity and industry leads to the energy crisis. The major energy crisis segment has been resolved through exploitation of fossil fuels. The com of burning of fossil fuel causes emission of greenhouse gases include carbon dioxide. Carbon emission from burned fuels causes climatic changes and global warming ~\cite{Fu2024, Kannan2016, Hanif2019}. As far as clean and renewable, sustainable energy is concerned. The conversion of light to chemical energy is a feasible method compared with the efficiency of energy converted from solar cells. For the first time, Honda and Fujishima., et al \cite{Fujishima1972}~explored the artificial photosynthesis of \ch{TiO2}. Absorption of photons in a semiconducting bandgap produces electromotive force in water, to split hydrogen and oxygen. Several wide semiconducting bandgap materials \ch{(TiO2,~ ZnO, ~WO3,~ Cds,~ CdSe,~ and~ Bi2Te3)} satisfy with the conduction band being more negative than hydrogen reduction potential and the valence band more positive than water oxidation potential. However, most of the transition metal oxides offer wide bandgap and absorb 4\% less ultraviolet light from the solar beam.
    
    As far as a narrow band gap is concerned, several transition metal tellurides such as MgTe, MnTe, TiTe, ZnTe, and \ch{Bi2Te3}~\cite{Negedu2022,Lu2023,PuthirathBalan2018,Bhat2017,DGao2021,Zhang2020}. The lack of tuning structural and electronic properties in single anionic compounds, material researchers moved to heteroanionic (mixed-anionic) compounds~\cite{Tripathi2021}. In other hand, excess heat from the sun to electrical energy in waste heat recovery or thermoelectric (TE) generators being a prominent method of sustainable energy harvest approach. To satisfy the energy demand of large scale industrial demands, to increase the conversion efficiency (ZT) of TEs. Typically, TE conversion efficiency of a material depends upon their high Seebeck coefficient, electrical conductivity, low thermal conductivity~\cite{Yin2019}. Metal tellurides, chalcogineds, half husler alloys, and zintl phases being an benchmark material. Subsequently, single anionic material inadequate with structurally tunable thermal properties~\cite{Ebling2007}.  
\\
\\
 Essentially absorption coefficient, electronic characteristics, defect tolerance, and carrier mobility in transition metal chalcogenides to compete. A band gap of 2.2 eV is necessary to achieve the desirable efficiency of over 15\%. As a cost-effective concern, transition metal-based compounds exhibit greater attention to optoelectronic applications. Recently, transition metal ions have been attracted to fluorescent active centers. In particular, pyrovanadate and chalcogenide clusters \ch{(VO4)} were attracted due to self-activated photoluminescence among several Vanadates \cite {Schira2020,Wang2022}.  One strategy is to enlarge the bandgap with multiple anions with \ch{d^5 (V^5^+, ~Nb^5^+,~ and~ Ta^5^+)} with s$^2$ (\ch{Ca^2^+, ~Sr^2^+, ~and~ Ba^2+}) \cite{He2020}. Group-II \ch{(Ca, Sr, ~and~ Ba)} and Group-IV \ch{(V, ~Ta,~ and~ Nb)} elements are promising candidates for the wide range of industrial development \cite{Jin2022}. However, several heteroanionic oxychalcogenide materials perform in seamless applications owing to their physical properties, such as photocatalysis, thermoelectric, superconductivity, nonlinear optics, etc~\citep{He2020,Kageyama2018, Mizuguchi2013, Pilania2020, Zhao2010, Ye2014, Duan2009}. 
\\
\\
	In general, anionic ordering and more covalent progression in heteroanionic materials offers layered chain-like structures with asymmetric coordination and non-centrosymmetry. Recently, Konatham, S., et al \cite{Konatham2023} experimentally reported several alkali-based heteroanionic materials. Prospect to the cost-effective energy consumption, alkali earth-based heteroanionic oxychalcogenides are earth-abundant and have excellent optoelectronic properties.  \ch{CaV2TeO8} has been identified as an optically active material with a bandgap of 2.16 eV which categories under a visible light region. At present, we utilized the first principles calculation to recognize the ground state properties of \ch{CaV2TeO8} heteroanionic oxychalcogenide such as structural, mechanical, electronic, optical, transport and thermoelectric properties towards energy-based applications. 
		\section{Computational Details}
To investigate the ground state properties of CaV$_2$TeO$_8$, first-principles implemented density functional theory (DFT) calculations (DFT) were performed. Full potential linear augmented plane wave (FP-LAPW) method WIEN2k package  was utilized \citep{Blaha2020} to solve the ground state equations. A generalized gradient approximation (GGA) with the Perdew-Burke-Ernzerhof (PBE) functional \citep{Perdew1996} was employed to fit the Birch-Murnaghan equation of state \citep{Murnaghan1937}. Furthermore, the electronic and optical properties of the material were calculated using the PBE exchange-correlation functional. The plane wave cutoff parameter was set as $RK_{\text{max}} = 7$; $G_{\text{max}} = 12 \, (\text{a.u.})^{-1}$; $l_{\text{max}} = 10$. The electronic configurations of the valence states were implemented as Ca [3s$^2$ 3p$^6$ 4s$^2$], V [3s$^2$ 3p$^6$ 3d$^3$ 4s$^2$], Te [4d$^{10}$ 5s$^2$ 5p$^4$],and O [2s$^2$ 2p$^4$]. The self-consistent-field threshold for iteration energy convergence was set to $10^{-6}$ Ry. Further, the Hellmann-Feynman forces on each atom were applied with a threshold of 0.01 eV/$\mathring{A}$. The high-symmetry Brillouin zone was sampled with an 6$\times$6$\times$2 $k$-point mesh to compute the ground state properties. The optimized crystallographic information is shown in the figure using the Visualization for Electronic and Structural Analysis software (VESTA) \citep{Momma2011}. The effective mass was calculated from the second-order polynomial fit of the band edges. Optical properties were calculated with 1500 $k$-points. Transport properties were investigated using the BoltzTrap code~\cite{MADSEN2006}. To estimate the charge carrier mobility in the bulk structure, Borden and Shockley \cite{Bardeen1950} derived an equation using the deformation potential theory\citep{Wan2023}, \citep{Gao2021}. The carrier mobility of CaV$_2$TeO$_8$ is given by the following equation \ref{eq:1}:
	
\[
\mu_{\text{CaV}_2\text{TeO}_8} = \frac{2\sqrt{2\pi} \, e\hbar^4 C_{3D}}{3 \left(k_B T\right)^{3/2} m^{*5/2} E_\text{ij}^2}
\label{eq:1}\tag{1}
\]
Here $k_B$ represents the Boltzmann constant, $\hbar$ is Planck’s constant, $T$ is the temperature (300 K), and $C_{3D}$ is the stretching in the unit cell. $C_{3D}$ is obtained from the elastic constant at the equilibrium volume, defined as
	
	\[
	C_{3D} = \frac{\left(\frac{\partial^2 E}{\partial \delta^2}\right)}{V_0}
	\tag{2}
	\]
	
	Here $E$ denotes the total energy due to the applied strain $\delta$, and $V_0$ represents the equilibrium volume of the unit cell. Further, $E_{ij}$ denotes the deformation potential, given by
	
	\[
	E_{ij} = \frac{\Delta E}{\Delta L / L_0}
	\tag{3}
	\]
	
	where $\Delta E$ represents the change in VBM or CBM energy versus lattice dilation (0.02\%). The effective mass of the charge carrier is denoted by $m^*$.
	
	\section{RESULTS and DISCUSSIONS}
	\subsection{Structural Properties}
	\subsubsection {Crystal Structure}
	
\begin{figure}[htbp]
		\centering
		\includegraphics[width=3.8 in]{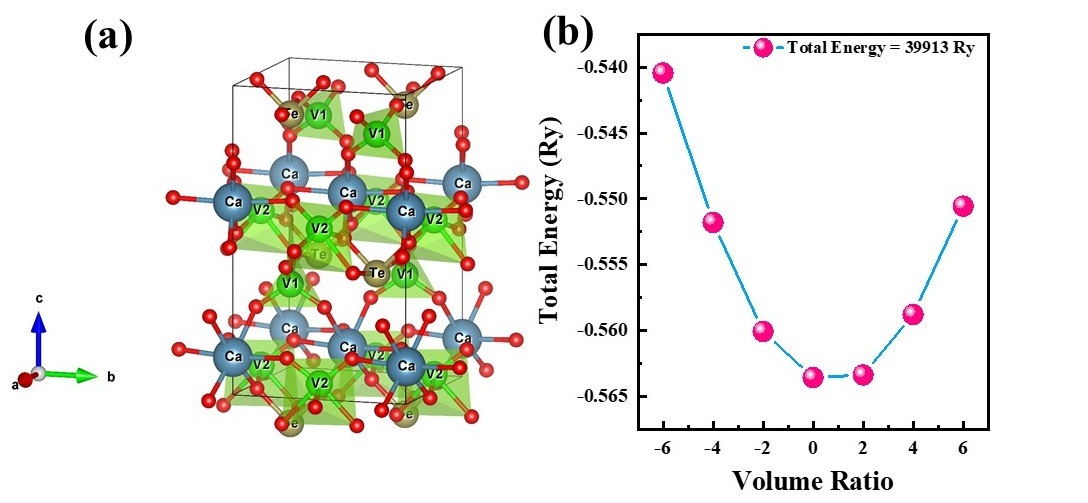}
		\vspace{-10pt}
		\caption{(a) Crystallographic ball-stick representation of CaV$_2$TeO$_8$. Here red, golden brown, green and blue balls are oxygen, tellurium, vanadium, and calcium atom respectively. (b) Volume variation with respect to energy diagram through the Mornaghan’s equation of state.}
		\label{fig:Crystal}
	\end{figure}

	Calcium vanadium tellurite belongs to the family of AB$_2$XO$_8$ heteroanionic oxychalcogenides, where A and B represent calcium and vanadium, respectively, bonded ionically to oxytellurite. Figure~\ref{fig:Crystal}(a) shows the crystal structure of CaV$_2$TeO$_8$, which crystallizes in the orthorhombic space group \textit{Ccc2}. Calcium and vanadium exhibit octahedral coordination with oxygen atoms, while a second vanadium atom forms a distorted tetrahedral coordination with oxygen. Furthermore, a telluride distorted tetrahedral bond with oxygen is observed. The optimized crystallographic information of CaV$_2$TeO$_8$ is listed in Table \ref{tab:structure}. Fig.~\ref{fig:Crystal} (b) shows the Total Energy versus c/a ratio. The crystallographic optimization of CaV$_2$TeO$_8$ was carried out through Murnaghan's equation of state (1). Here $E_0$ denotes the ground state energy (T = 0K), $B_0$ and $B_0^{\prime}$ stand for the bulk modulus and the pressure derivative of the bulk modulus. $V$ is the volume of the unit cell and $V_0$ is the equilibrium volume, respectively.
	
	\[
	E(V) = E_0 + \left[ \frac{V_0}{V} \right]^{B_0^{\prime}} \left( \frac{1}{B_0^{\prime} - 1} + 1 \right) - \frac{B_0 V_0}{B_0 - 1} \tag{4}
	\]
		\begin{table}[h!]
		\caption{\label{tab:structure}Structural parameters of CaV$_2$TeO$_8$ in the \textit{Ccc2} space group. Lattice constants are in \AA, and the unit cell volume is in \AA$^3$.}
		\begin{ruledtabular}
			\begin{tabular}{cccc}
				Space Group & Lattice Constants (\AA) & & Volume (\AA$^3$)\\
				\hline
				\textit{Ccc2} & $a=7.0796$, $b=7.2356$, $c=12.4446$ & & 1138.235 \\
\end{tabular}
	\end{ruledtabular}
\end{table}

	\subsubsection {Mechanical Properties}
	It is crucial to investigate the mechanical stability of the material towards energy-related applications. Fundamentally, the elastic constant ($C_{ij}$) is a proportionality between an applied lattice strain on the crystal lattice and the energy variation due to lattice strain. The elastic constants of a material are distinguished concerning their crystal systems. Currently, CaV$_2$TeO$_8$ belongs to the orthorhombic crystal system. The orthorhombic system has lower symmetry and nine fundamental elastic constants: $C_{11}$, $C_{12}$, $C_{13}$, $C_{22}$, $C_{23}$, $C_{33}$, $C_{44}$, $C_{55}$, and $C_{66}$. The observed elastic constant values are shown in Table \ref{table:elastic_constants} and Table \ref{table:elastic_wave_parameters}.
	\begin{table}[ht]
		\caption{Elastic constants $C_{ij}$ (in GPa) for the material.}
		\begin{ruledtabular}
			\begin{tabular}{ccccccccc}
				\text{$C_{11}$} & 
				\text{$C_{12}$} & 
				\text{$C_{13}$} & 
				\text{$C_{22}$} & 
				\text{$C_{23}$} & 
				\text{$C_{33}$} & 
				\text{$C_{44}$} & 
				\text{$C_{55}$} & 
				\text{$C_{66}$} \\ 
				\hline
				302.80 & 72.36 & 92.84 & 339.48 & 141.26 & 210.37 & 137.29 & 70.41 & 62.56 \\
			\end{tabular}
		\end{ruledtabular}
		\label{table:elastic_constants}
	\end{table}
	The elastic stability of the orthorhombic crystal system was derived from Born Stability criteria. The Born criteria at 0 GPa satisfy the elastic stability of the orthorhombic system \citep{Lau1998}:
	
	\[
	(C_{11} + C_{22} - 2C_{12}) > 0 \tag{5}
	\]
	
	\[
	(C_{11} + C_{33} - 2C_{13}) > 0 \tag{6}
	\]
	
	\[
	(C_{22} + C_{33} - 2C_{23}) > 0 \tag{7}
	\]
	
	\[
	C_{ij} > 0 \label{eq:8}\tag{8}
	\]
	
	\[
	(C_{11} + C_{22} + C_{33} + 2C_{12} + 2C_{13} + 2C_{23}) > 0 \tag{9}
	\]
	
 Further, we utilized the Voigt-Reuss-Hill method to derive elastic properties. The calculated elastic parameters are listed in Table \ref{table:elastic_constants}.In CaV$_2$TeO$_8$, the bulk modulus was higher to compare to the shear modulus, suggesting higher resistance to change in volume than to shape distortion. Furthermore, the ratio between Young's and shear moduli depicts Pugh’s ratio. The calculated Pugh’s ratio (1.80) and Poisson’s ratio (0.26) suggest the material is ductile. Moreover, the transverse ($v_t$) and longitudinal ($v_l$) velocities of CaV$_2$TeO$_8$ are listed in Table \ref{table:elastic_wave_parameters}.
\begin{figure}[htbp]
	\centering
	\includegraphics[width=3.4in]{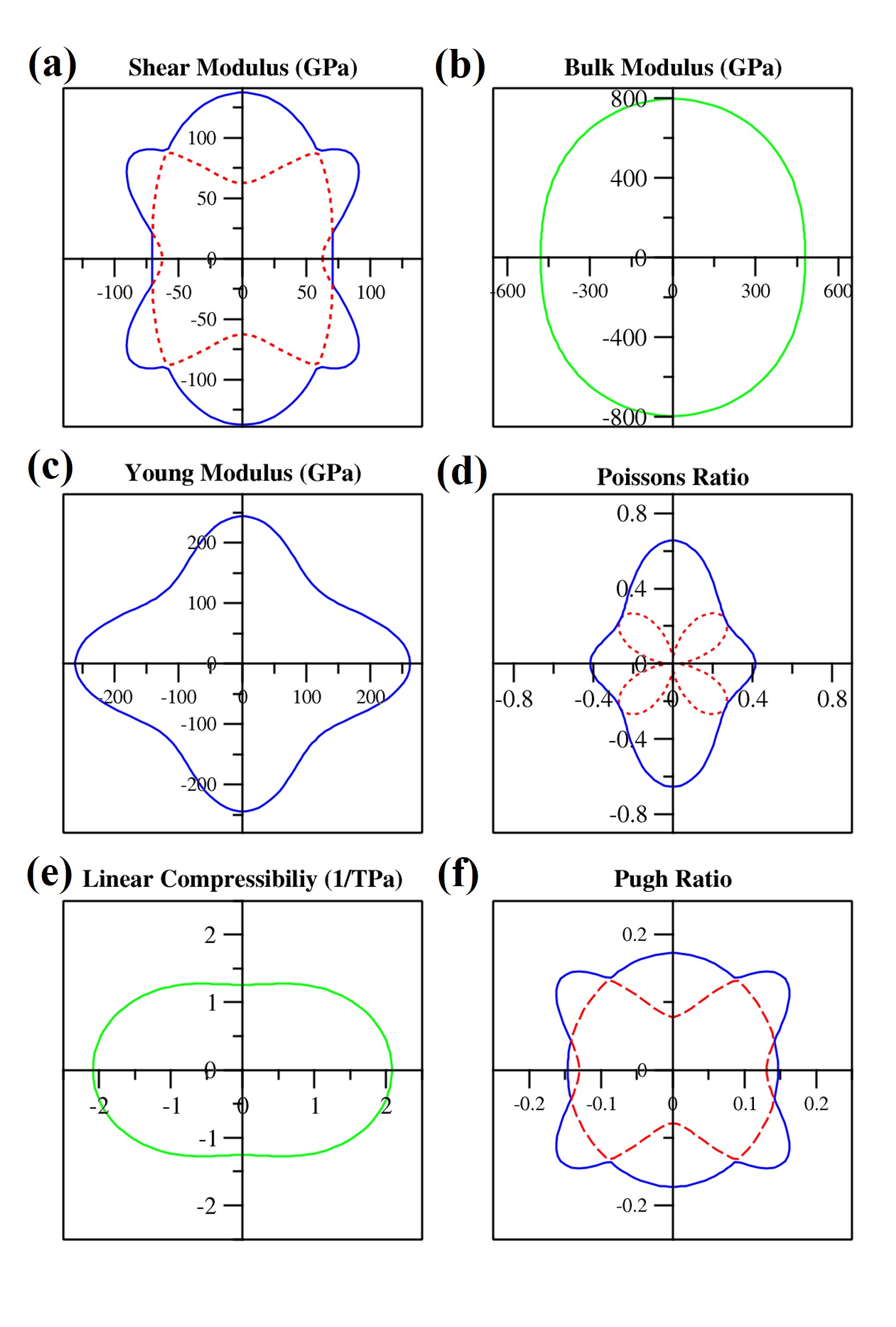}
	\caption{Mechanical Anisotropy of CaV$_2$TeO$_8$ within orthorombic [100] plane. (a) Shear Modulus (GPa), (b) Bulk Modulus, (c) Youngs Modulus, (d) Poissons Ratio, (e) Linear Compressibility (1/TPa), and (f) Pugh Ratio.}
	\label{fig:Anisotropy}
\end{figure}
	
	\begin{table}[ht]
		\caption{Elastic wave velocities, Acoustic Debye temperature, and Grüneisen parameter.}
		\begin{ruledtabular}
			\scalebox{0.8}{ 
				\begin{tabular}{cccccc}
					\text{($\vartheta_l$) (m/s)} & 
					\text{($\vartheta_t$) (m/s)} & 
					\text{($\vartheta$) (m/s)} & 
					\text{($\Theta_D$) (K)} & 
					\text{($\gamma$)} \\ 
					\hline
					8119.9 & 4517.77 & 9681.10 & 632.509 & 1.63 \\
				\end{tabular}
			}
		\end{ruledtabular}
		\label{table:elastic_wave_parameters}
	\end{table}
	 Debye temperature is an essential factor in several physical properties such as bond nature, specific heat, and thermal expansion. The calculated elastic Debye temperature (632.5 K) suggests that CaV$_2$TeO$_8$ has high lattice thermal conductivity. However, the orthorhombic crystalline planes are axially dependent on other planes, as determined by Zener’s Anisotropic index. The mechanical isotropic index is determined as $A = 1$. On the other hand, anisotropy is defined as $A \neq 1$. Further, the directional dependency of crystal plane [100] and (010) mechanical anisotropy characteristics of CaV$_2$TeO$_8$ can be derived \citep{Lau1998} from the following relation\ref{eq:10} :
	
	\[
	A = \frac{2C_{55}}{C_{33} - C_{13}} \label{eq:10} \tag{10}
	\]
	
	We investigated the directional dependence of the elastic properties with the ElATool code~\cite{Liu2022}. We obtained the Zener’s anisotropic factor as 1.19 for the reference axis of [001] and symmetry plane (010) direction. Fig.~\ref{fig:Anisotropy} illustrates the 2D representation of (a) shear modulus, (b) bulk modulus, (c) Young’s modulus, (d) Poisson’s ratio, (e) linear compression and (f) Pugh ratio along the directional (100) plane. In the present study, the overall mechanical properties of CaV$_2$TeO$_8$ suggest that the material is mechanically stable and a promising candidate for flexible photovoltaic and optoelectronic devices.
	\subsection{Electronic Structure}
	\subsubsection{Band Structure Properties}
\begin{figure}[htbp]
	\centering
	\includegraphics[width=3.25in]{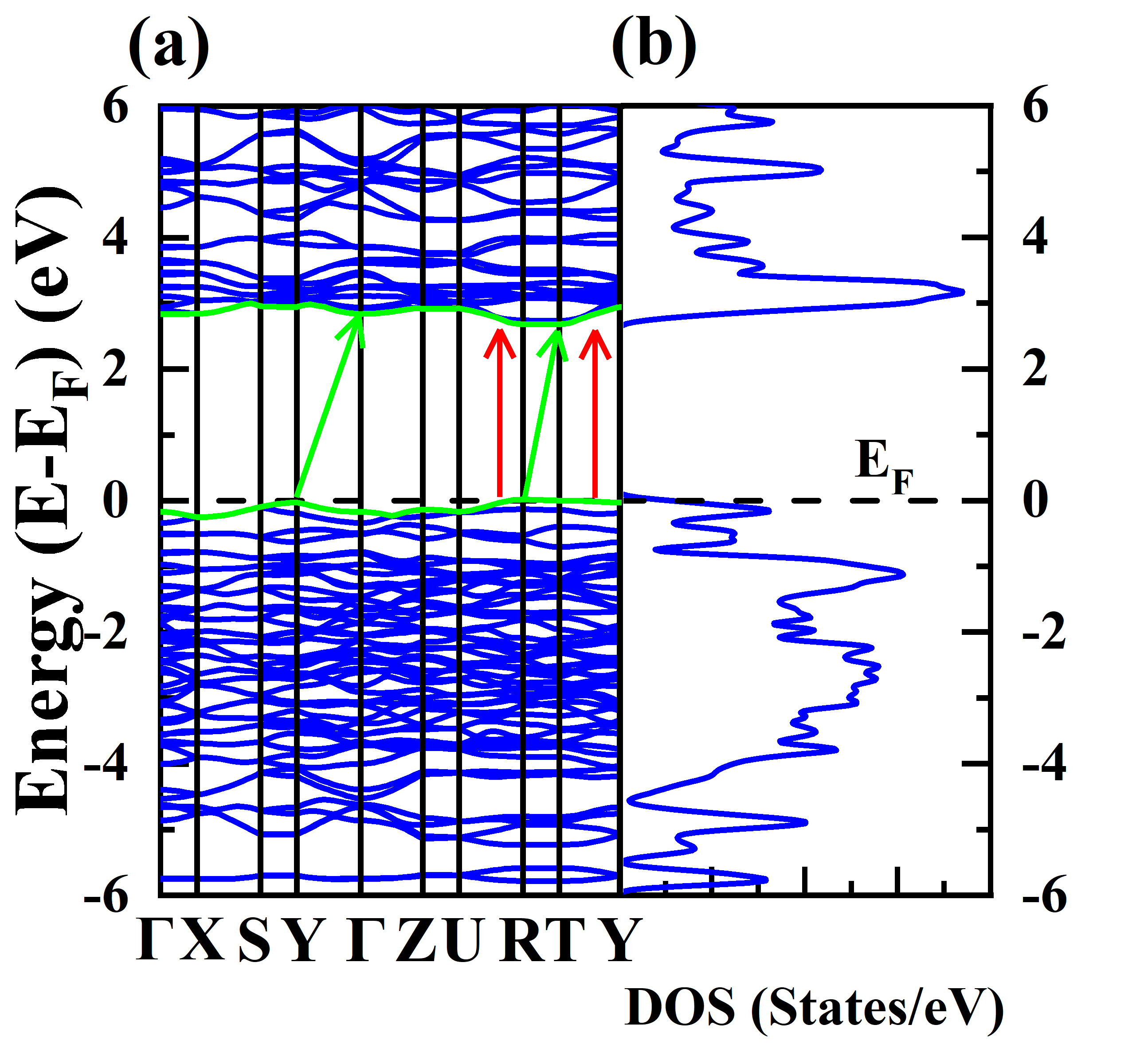}
	\caption{(a) Electronic band structure of CaV$_2$TeO$_8$ with the Fermi energy adjusted to zero. (b) Total density of states (DOS) for CaV$_2$TeO$_8$.}
	\label{fig:band_structure}
\end{figure}

To understand the electronic properties such as density of states (DOS) and band structure calculations were performed. Fig.~\ref{fig:band_structure} (a) shows CaV$_2$TeO$_8$ the electronic band structure, fermi level adjusted to zero. A parallel plot in Fig.~\ref{fig:band_structure} (b) shows the total density of states (DOS) for CaV$_2$TeO$_8$. The energy bands below the Fermi level correspond to the valence band, while the bands above the Fermi level represent the conduction band. Furthermore, the valence band maximum (VBM) is located at 0 eV, and the conduction band minimum (CBM) is observed at 2.86 eV. The electronic band structure is plotted as a function of the orthorhombic high-symmetry Brillouin zone points, with the $k$-path given by ($\Gamma$-X-S-Y-$\Gamma$-Z-U-R-T-Y). The conduction band minimum is found at the X, R points, while the valence band maximum occurs between the Y, R, and T high-symmetry points. The electronic bandgap of CaV$_2$TeO$_8$ is observed to be 2.86 eV, with both direct (R-R) and (T-T) transitions, as well as indirect band transitions (R-T), (Y-X) at high-symmetry points. Additionally, the total density of states plot from Fig.~\ref{fig:band_structure} (b) confirms the band opening of 2.86 eV using the GGA exchange-correlation functional. The observed bandgap marginally deviates from the experimentally a tau's plot reported value of 2.16 eV \cite{Konatham2023}.
Further, compressive and tensile strain of -6 to 6\% were subjected to a and c biaxial. For negative strain values (e.g., -6\%, -4\%), the bandgap decreases from 2.67 eV to 2.65 eV. This suggests that compressive strain causes the atomic spacing to reduce, potentially increasing the overlap of electronic orbitals, which leads to a narrowing of the bandgap.
This behavior is typical in semiconductors where compressive strain tends to lower the energy required to transition an electron from the valence band to the conduction band. similarly, For positive strain values (e.g., 2\%, 4\%, 6\%), the bandgap similarly decreases (from 2.80 eV at 0\% strain to 2.65 eV at 6\%). Tensile strain causes the atoms to be pulled further apart, which may lower the interaction between electronic states, reducing the bandgap.
\begin{figure}
	\centering	\includegraphics[width=0.5\textwidth]{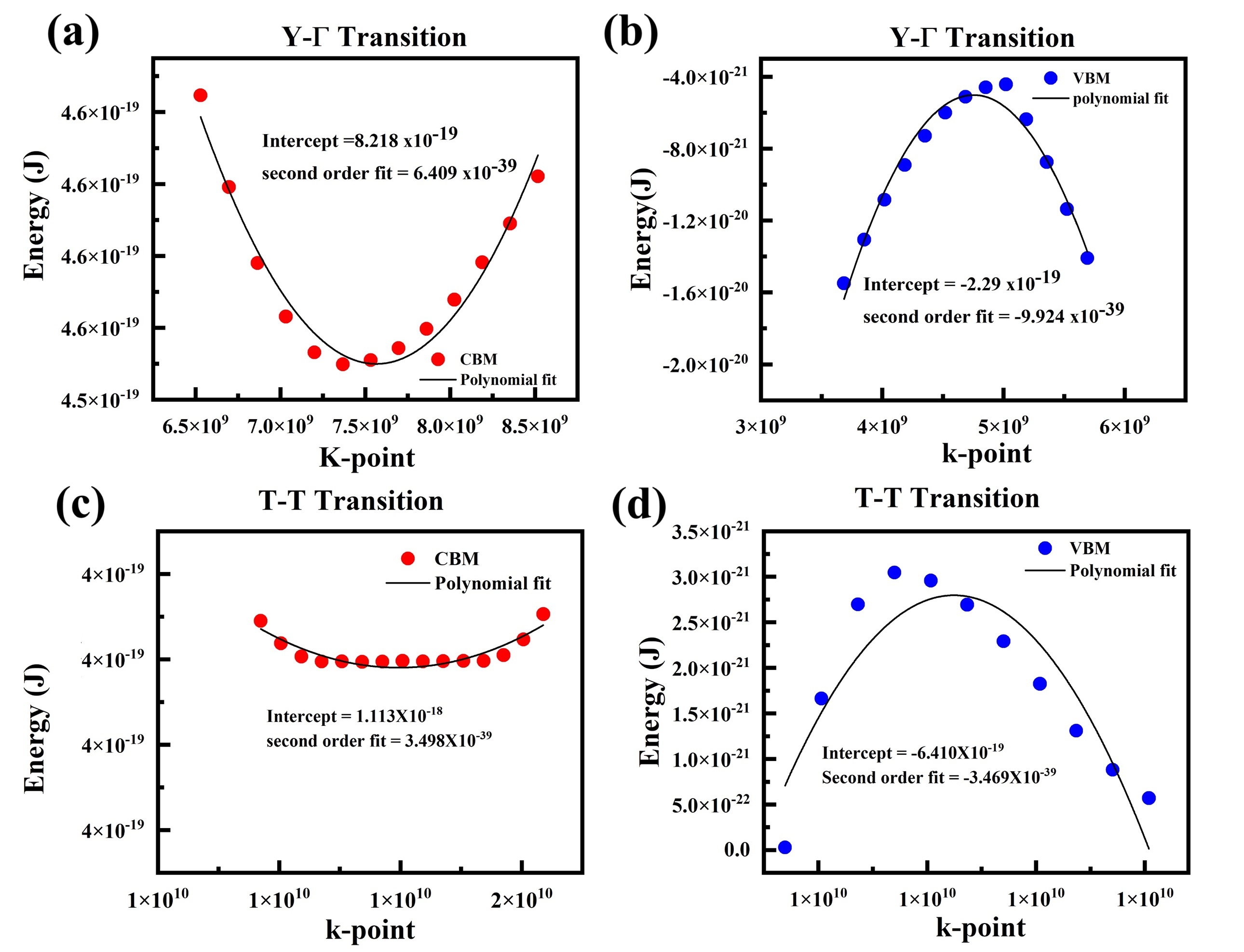} 
	\caption{Energy transitions for (a) Y$-$Γ Transition (CBM), (b) Y$-$Γ Transition (VBM), (c) T$-$T Transition (CBM), and (d) T$-$T Transition (VBM). The polynomial fit and intercept values are indicated for each case.}
	\label{fig:supplementary}
\end{figure}
\begin{table}
	\caption{Effective masses of electron and hole, relative mass, dielectric constant, exciton binding energy, and bohr radius for different transitions.}
	\begin{center}
		\begin{ruledtabular}
			\begin{tabular}{rcccccccc}
				\text{Transition} & 
				\text{$m_e^*$} & 
				\text{$m_h^*$} & 
				\text{R} & 
				\text{$m_\mu$} & 
				\textepsilon$_\parallel$ $\times$ \textepsilon$_\perp$ & 
				\text{$E_\text{ex}$} & 
				\text{R$_B$} \\ & & & & ($m_0$)& &(meV) & \text{(\AA)} \\ 
				\hline
				T-T & 1.74 & 1.73 & 0.99 & 0.86 & 6.5 & 1812.3 & 1.26 \\ 
				Y-$\Gamma$ & 0.65 & 0.94 & 1.44 & 0.38 & --- & 795.41 & 3.75 \\
			\end{tabular}
		\end{ruledtabular}
	\end{center}
	\label{tab:exciton_properties}
\end{table}
The investigated results of the band structure calculation indicate an indirect electronic transition with a bandgap of 2.86 eV. Consequently, the active electrons in the conduction band exhibit an unfeasible direct transition to the valence band. Instead, the transfer of electrons within the indirect band requires additional photons. In principles, extendeing the lifetime of photoinduced electrons compared to direct band transitions results in increased difussion length and prolong reaction for carriers. Furthermore, the recombination rate of photogenerated carriers plays a vital role in determining photocatalytic performance. In general, the carrier recombination rate is associated with the relative effective mass of the hole and electron, expressed as $R = \frac{m_h^*}{m_e^*}$. A higher value of the relative effective mass ratio ($R$) corresponds to a lower recombination rate, which can enhance photocatalytic efficiency~\cite{Zhang2015}~\cite{VIJAY2022}.
The obtained $R$-value is approximately 0.99 and 1.44 for the T-T and Y-$\Gamma$ transition respectively in CaV$_2$TeO$_8$, indicating that the separation of photogenerated carriers can be efficient within the photocatalytic reaction. Furthermore, the binding energy of excitons were computed using equation \eqref{eq:11}:
\begin{equation}
	E_{\text{ex}} = \frac{m_{\mu} R_{\infty}}{m_0 \epsilon_0^2} \tag{11}  \label{eq:11}
\end{equation}
where $1/m_{\mu} = \left( \frac{1}{m_e^*} + \frac{1}{m_h^*} \right)$ represents the reduced effective mass, $R_{\infty}$ is the Rydberg constant, $m_0$ is the rest mass of an electron, and $\epsilon_0^2$ stands for the dielectric constant along the parallel and perpendicular axes (\textepsilon$_\parallel$ $\times$ \textepsilon$_\perp$) along with the $z$ direction. The binding energy of the photogenerated carriers is listed in Table~\ref{tab:exciton_properties}. Direct and indirect transitions exhibit binding energies of 1812.3 meV and 795.41 meV, respectively, required to separate the photogenerated fermions. 

The curved nature of CBM and VBM indicates a lower effective mass of charge carriers in the Y-$\Gamma$ transition. Conversely, the flat band of the T-T direct transition between CBM and VBM shows larger effective masses. The lower effective masses of the photogenerated charge carriers in the Y-$\Gamma$ transition suggest higher carrier mobility compared to the T-T transition. Moreover, the excitonic radius of charge carriers can be characterized into two types. They are Mott-Wannier and Frenkel excitons, in comparison to the optimized lattice constant of CaV$_2$TeO$_8$. The excitonic radius is investigated from the relation:
\begin{equation}
	R_B = \frac{\varepsilon_1(0) \times m_0}{m_{\mu}} R_0 \tag{12}
\end{equation}

The calculated Bohr radii for the direct and indirect transitions were 1.26~Å and 3.75~Å, respectively. The Bohr radius of the excitons is smaller compared to the crystallographic lattice parameter. A smaller radius of excitons suggests that they are of the Frenkel type, or strongly bound excitons. where $R_B$ is the Bohr radius (0.529~Å). To investigate the photocatalytic mechanism of CaV$_2$TeO$_8$ material's redox potentials for the CB and VB edges at zero-point charge of 0.186 eV (0.186~eV) were calculated \cite{Nayak2021}. The conduction and valence band potential edges, $E_{\text{CB}}$ and $E_{\text{VB}}$, are calculated using the following equations ~\eqref{eq:13} and ~\eqref{eq:14}:
\begin{equation}
E_{\text{CB}} = \chi - E_C - \frac{E_g}{2} \label{eq:13} \tag{13}
\end{equation}
\begin{equation}
	E_{\text{VB}} = E_{\text{CB}} + E_g \quad \label{eq:14} \tag{14}
\end{equation}

The electronegativity $\chi$ is calculated using the formula \eqref{eq:15}:

\begin{equation}
	\chi = {^{12}\sqrt{\chi_{\text{Ca}}^1 \cdot \chi_{\text{V}}^2 \cdot \chi_{\text{Te}}^1 \cdot \chi_{\text{O}}^8} } \label{eq:15} \tag{15}
\end{equation}

where $E_C$ is the free energy corresponding to the hydrogen evolution reaction (4.5~eV), $E_g$ is the bandgap, and $\chi$ is the absolute electronegativity of CaV$_2$TeO$_8$ obtained from Mulliken analysis. The calculated bandgap of the material is 2.8~eV for the direct band transition.
The electronegativity $\chi$ of a semiconducting material is computed as the arithmetic mean of the first ionization potential and atomic electron affinity. The electronegativity ($\chi$) of CaV$_2$TeO$_8$ was found to be 5.98~eV. The symbolic representations of $\chi_{\text{Ca}}$, $\chi_{\text{V}}$, $\chi_{\text{Te}}$, and $\chi_{\text{O}}$ refer to the individual electronegativities of the metal and oxide atoms. Furthermore, the conduction band edge ($E_{\text{CB}}$) and valence band edge ($E_{\text{VB}}$) at point zero charge are calculated to be 0.12~eV and 3.08~eV (vs NHE), respectively.
\begin{figure}[htbp]
	\centering
	\includegraphics[width=3.25in]{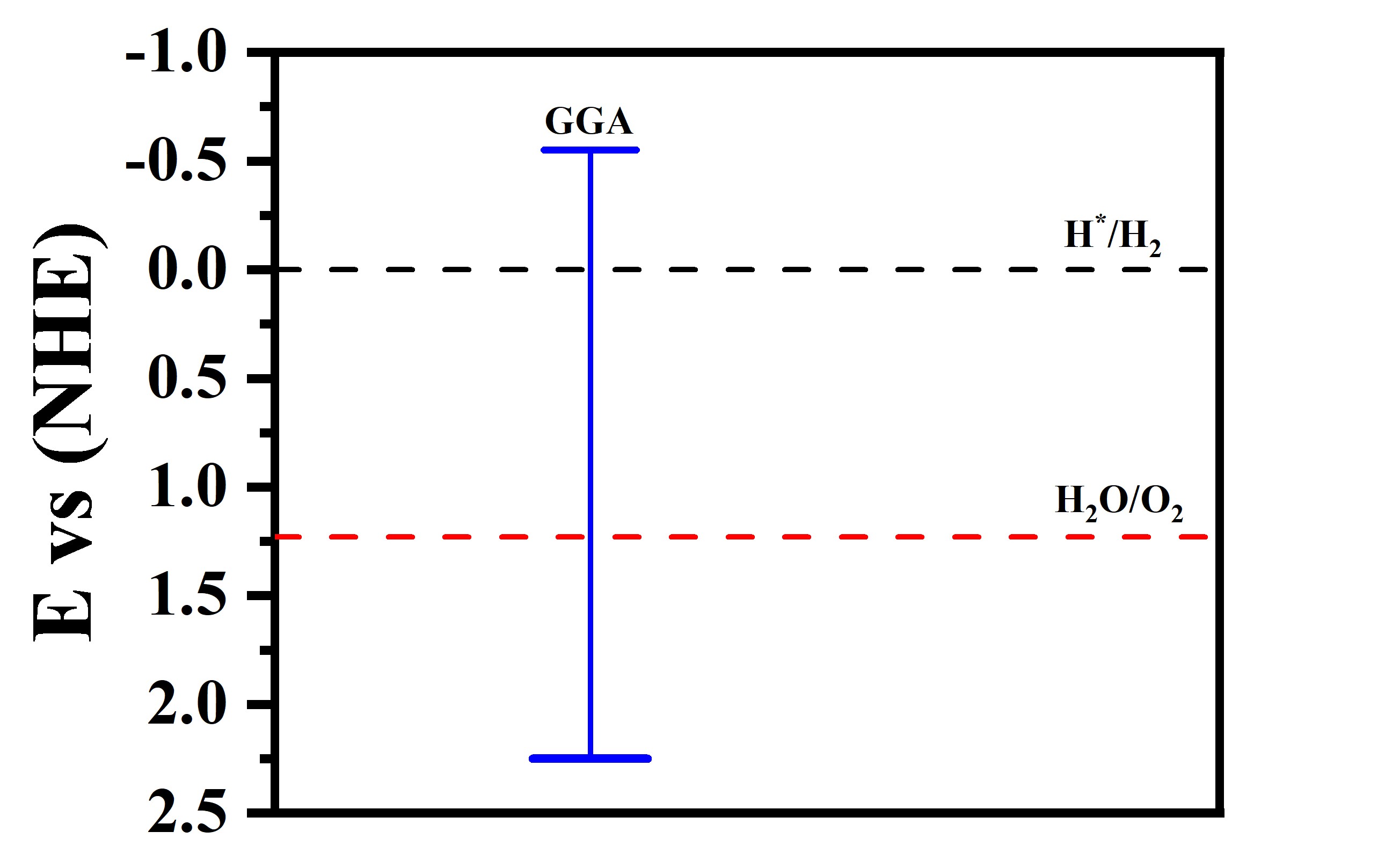}
	\includegraphics[width=3.25in]{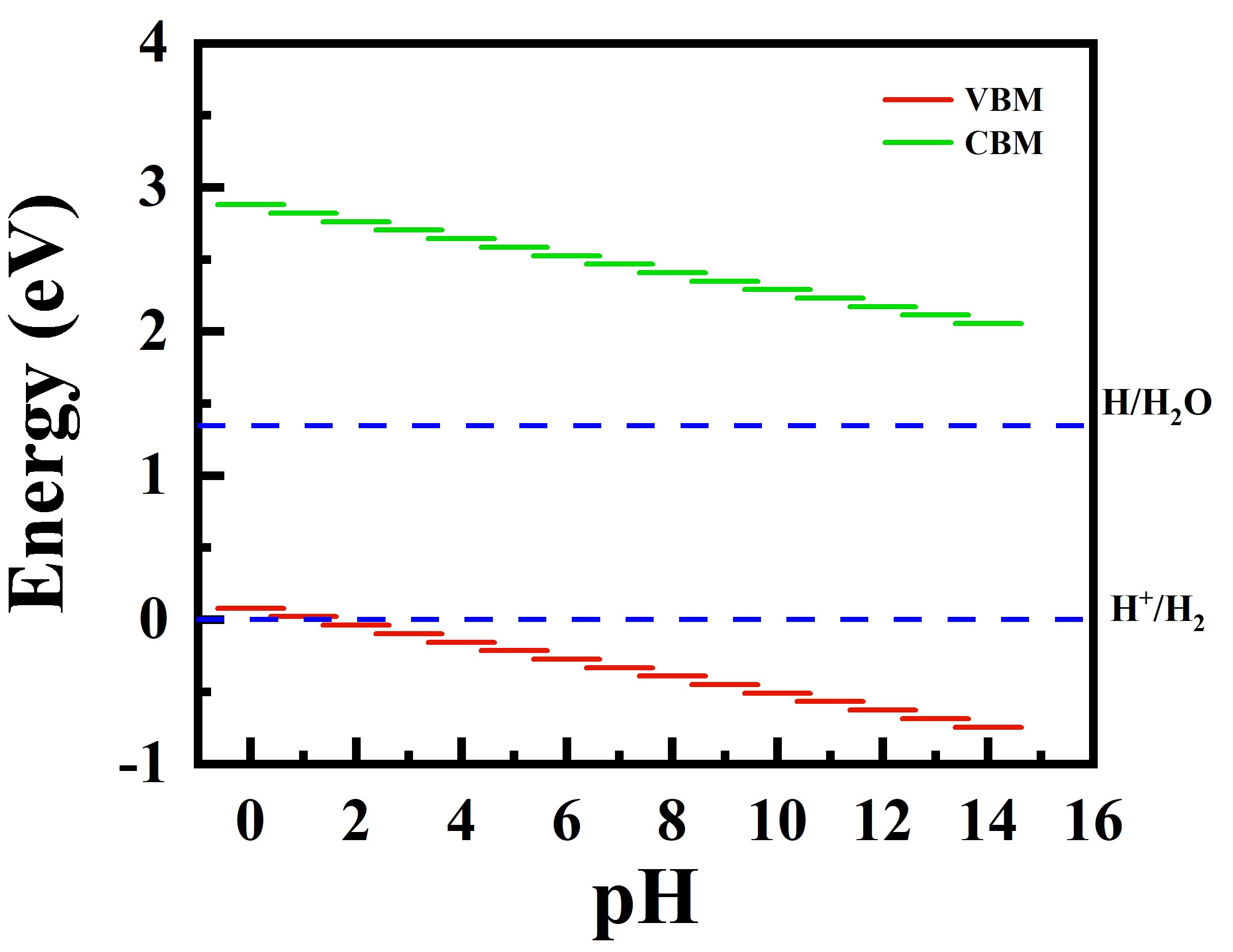}
	\vspace{-15pt}
	\caption{ Photocatalytic hydrogen evolution VBM and CBM potential window for CaV$_2$TeO$_8$ from pH 0 to 12.}
	\label{fig:Band_edge_vs_pH}
\end{figure}

The illustrated figure Fig.~\ref{fig:Band_edge_vs_pH}represents the appropriate band edge position of CaV$_2$TeO$_8$ within the redox potentials of water. The change in pH value of the medium is a crucial factor for the photocatalytic reaction~\cite{Bardeen1950,Nayak2021}. In that aspect, oxidation and reduction potentials of water with pH values are obtained from the following equation ~\eqref{eq:16} and \eqref{eq:17}:

\begin{equation}
	E_{\text{O}} = -5.67 + 0.059 \times \text{pH} \quad \label{eq:16} \tag{16}
\end{equation}
\begin{equation}
	E_{\text{R}} = -4.44 + 0.059 \times \text{pH} \quad \label{eq:17} \tag{17}
\end{equation}

The conduction band minimum (CBM) position is higher than the water reduction potential (NHE vs vacuum) of -4.44 eV, and the valence band maximum (VBM) is lower than the water oxidation potential (NHE vs vacuum) of -5.67 eV. The Fig.~\ref{fig:Band_edge_vs_pH} illustrates the suitable position of the band edges with a pH range of zero.
\subsubsection{Density of States}
\begin{figure}[htbp]
	\centering
	\includegraphics[height=11cm, width=1\linewidth]{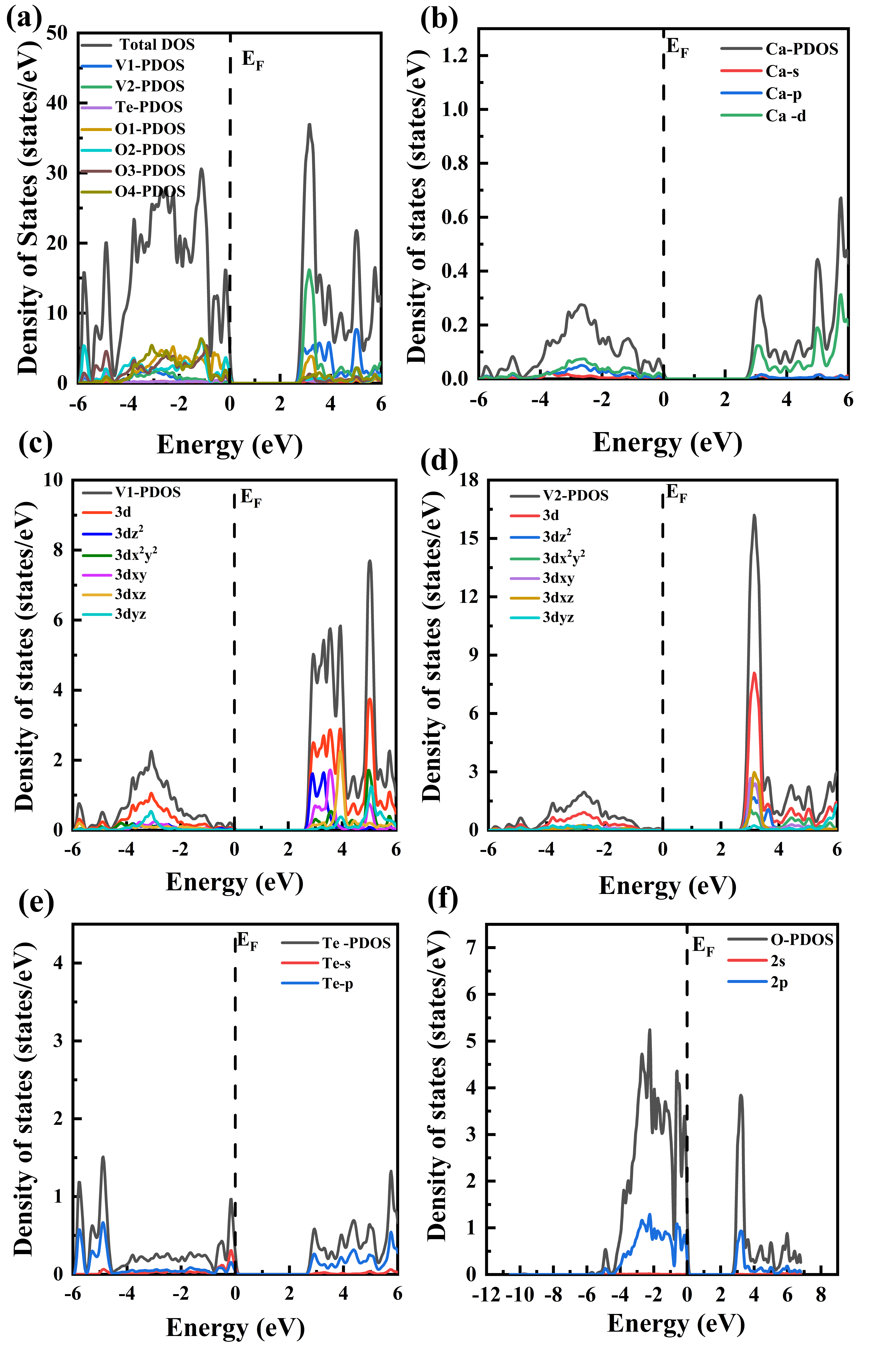}
	\vspace{-15pt}
	\caption{(a) Total and partial density of states (DOS) of CaV$_2$TeO$_8$. (b) Partial DOS of calcium. (c) Partial DOS for V1, (d) V2, (e) Te, and (f) O1.}
	\label{fig:dos}
\end{figure}

The total and partial density of states (TDOS\&PDOS) were calculated to understand the valence and conduction band contributions in the semiconducting band structure. The Fermi level is plotted at 0 eV to classify the valence and conduction bands based on atomic contributions. The TDOS and PDOS of Ca, V, Te, and O are shown in Fig.~\ref{fig:dos}. Fig.~\ref{fig:dos}(a) shows the valence and conduction band contributions of individual atoms present in the unit cell. The separation of the valence and conduction bands in the DOS exhibits a band gap of 2.8 eV. The total DOS profile mirrors the band alignment observed in the band structure plot, with the Fermi level drawn at zero energy in Fig.~\ref{fig:dos}(a). The Fermi energy level is split into two major regions: valence and conduction states. The valence and conduction states are located in the negative and positive energy regions, respectively. The vanadium (V), tellurium (Te), and oxygen (O) total DOS contribute significantly to both the valence and conduction states.

 The individual atomic orbital representations in Fig.~\ref{fig:dos}(b) show the calcium partial DOS, which makes a minor contribution to the valence and conduction states. In Fig.~\ref{fig:dos}(b), the valence and conduction states are distributed with p-like and d-like states. Conversely, the V, Te, and O partial density of states make a major contribution to the bandgap. Fig.~\ref{fig:dos}(c), (d), (e), and (f) represent the partial density of states for V1, V2, Te, and O1. The valence state below 0 eV shows a major doublet peak mimicking the p-like states in O and Te. Specifically, the localized $d_{t_{2g}}$ orbitals split into $d_{xy}$, $d_{xz}$, and $d_{yz}$ states. Similarly, the excited states of the $d_{eg}$ orbitals are classified as $d_{x^2-y^2}$ and $d_{z^2}$. Further, the $d$ state of V1 splits into two components in Fig.~\ref{fig:dos}(c), namely $t_{2g}$ and $e_g$ states. The V1-$3d$ orbital valence state contributes from the $t_{2g}$-$d_{yz}$ state. The conduction state of V1 shows a prominent contribution from the $3d$-$e_g$ states, namely $3d_{z^2}$, $3d_{xy}$, and $3d_{yz}$, which are represented in blue, orange, and sky blue colors, respectively. The splitting from the V1-$3d$ orbital indicates the sharing of the O-$p$ orbital. On the other hand, the V2-PDOS in Fig.~\ref{fig:dos}(d) shows that the conduction states belong to both $t_{2g}$ and $e_g$ states. From Fig.~\ref{fig:dos}(f), the conduction states of the O $p$-like state may be due to bonding with V1, V2, and Te atoms. The unique partial density of states from V1 and V2 shows octahedral and tetrahedral sharing with the oxygen atoms, respectively. The obtained DOS plots demonstrate electron transfer from non-centrosymmetric anionic layer of \([\mathrm{V_2TeO_8}]^{2-}\) and good agreement with the structural configuration of CaV$_2$TeO$_8$.

\subsection{Optical Properties and Dielectric Functions}

In principle, photocatalytic water splitting depends on optical properties such as absorption of the solar spectrum, dielectric constant, refraction, optical conductivity, and energy loss spectrum. The optical properties of CaV$_2$TeO$_8$ are attributed to the allowed electronic energy levels. In other words, the dielectric functions of CaV$_2$TeO$_8$ are the key parameters that influence the optical characteristics. As far as theoretical investigations of photocatalysis are concerned, the absorption of visible light is a crucial factor. The optical response of CaV$_2$TeO$_8$ is characterized by its frequency-dependent dielectric functions. From optical parameters, one can derive optical characteristics such as electron mobility and recombination rates. The essential dielectric function $\varepsilon(\omega)$ of optical characteristics is determined~\cite{Linear2006} from the following equation \ref{eq:18}:
\begin{equation}
	\varepsilon(\omega) = \varepsilon_r(\omega) + i \varepsilon_i(\omega) \tag{18} \quad \label{eq:18}
\end{equation}
where $\varepsilon_r(\omega)$ and $\varepsilon_i(\omega)$ denote the real and imaginary parts of the dielectric function, respectively. Typically, in semiconductors, intra-band transitions are emphasized by $\varepsilon_i(\omega)$ ~\citep{Linear2006}. Further, the two parts of the dielectric functions, $\varepsilon_r(\omega)$ and $\varepsilon_i(\omega)$, are obtained from the following equations \ref{eq:19} and \ref{eq:20}:
\begin{equation}
	\varepsilon_r(\omega) = 1 + \frac{2p}{\pi} \int_0^\infty \frac{\omega' \varepsilon_i(\omega')}{\omega'^2 - \omega^2} d\omega' \tag{19} \quad \label{eq:19}
\end{equation}
\begin{equation}
	\varepsilon_i(\omega) = \frac{8}{2\pi \omega^2} \sum_{n,n'} \int |P_{nn'}(k)|^2 \frac{dS_k}{\nabla \omega_{nn'}(k)} \tag{20} \quad \label{eq:20}
\end{equation}
Here, $P_{nn'}(k)$, $\nabla \omega_{nn'}(k)$, and $S_k$ represent the dipole matrix, energy difference, and surface area, respectively. The real part of the dielectric function, $\varepsilon_r(\omega)$, correlates with the polarization and dispersion. Conversely, the imaginary part of the dielectric function corresponds to energy loss spectra. The observed optical absorption at approximately 3 eV and the static dielectric function from the real part $\varepsilon_0(\omega)$ were calculated to be 2.1 using the PBE exchange-correlation functional. Three sharp peaks of the real part dielectric function~$\varepsilon_r(\omega)$ near 4.2 eV, 5.5 eV, and 7.1 eV in the UV region denote robust polarization. Eventually, the real part of the dielectric function $\varepsilon_r(\omega)$ tends to become negative as an increasing function of photon energy, approaching 10 eV.\\
 \begin{figure}[htbp]
	\centering
	\includegraphics[width=1\columnwidth]{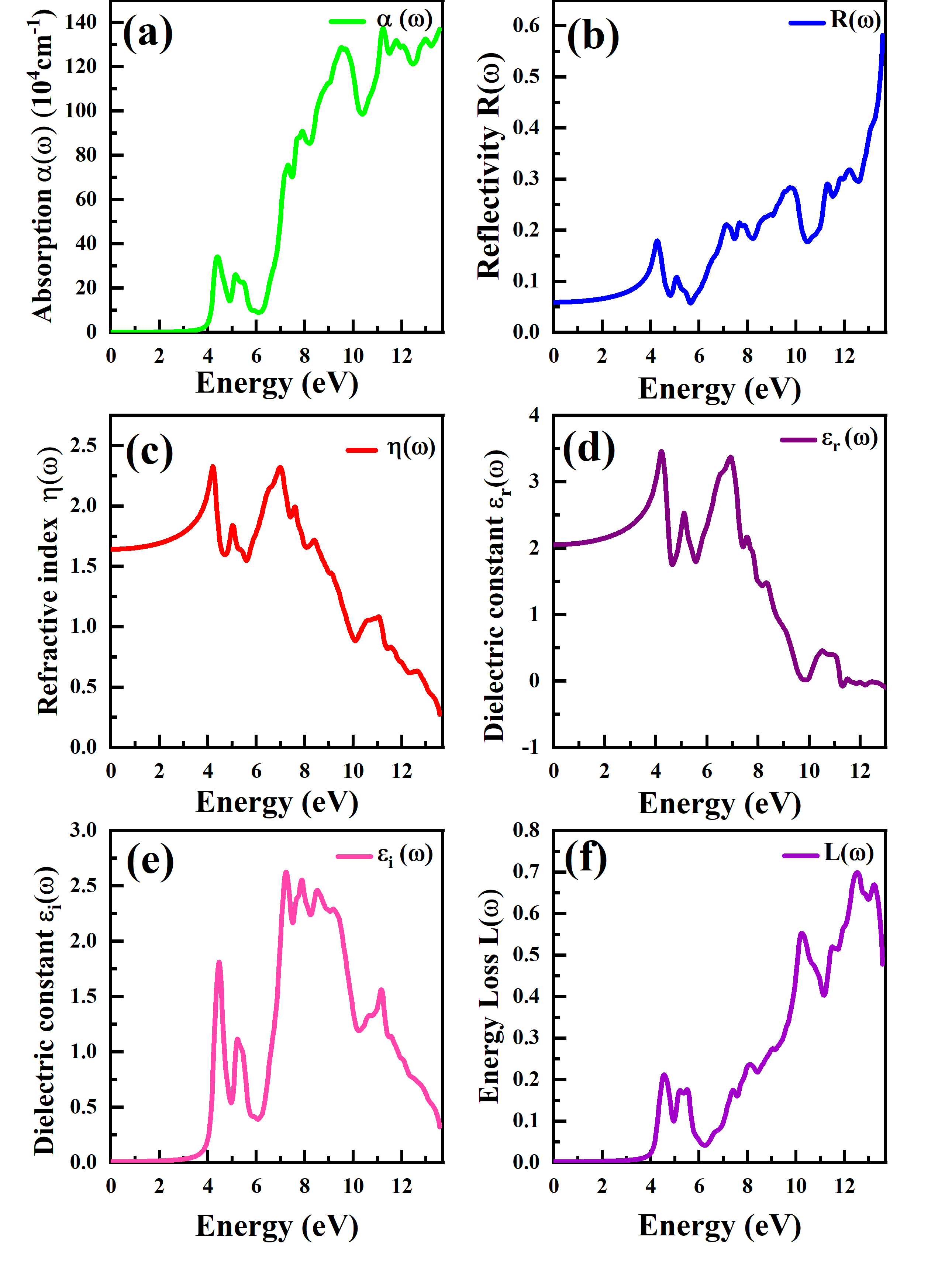}
	\caption{Optical properties of CaV$_2$TeO$_8$: (a) absorption, (b) reflectivity, (c) refractive index, (d) real, (e) imaginary components of dielectric constant, and (f) energy loss spectra.}
	\label{fig:optical_properties}
\end{figure}Further, the plotted imaginary part of the dielectric function $\varepsilon_i(\omega)$ as a function of photon energy is shown in the Fig.~\ref{fig:optical_properties}(d) and (e). The imaginary part of the dielectric constant is a suitable candidate to identify the material's absorption of incident photons within the absorption band. The first sharp peak at a photon energy of 4.2 eV with a peak value of 1.8 is observed. This sharp peak denotes the material’s absorption energy belonging to the UV region. To determine the specific energy region of the material, the optical absorption coefficient $\alpha(\omega)$ is an important parameter. It evaluates the amount and depth of light penetration within the material before absorption takes place. The absorption coefficient $\alpha(\omega)$ is defined~\cite{Linear2006} by the following relation \ref{eq:21}:
\begin{equation}
	\alpha(\omega) = \sqrt{2 \omega} \left[\sqrt{\varepsilon_r^2(\omega) + \varepsilon_i^2(\omega) - \varepsilon_r(\omega)}\right]^{1/2} \tag{21} \quad \label{eq:21}
\end{equation}
The calculated optical absorption for CaV$_2$TeO$_8$ is shown in Fig.~\ref{fig:optical_properties}(b). The maximum intensity of absorption is observed in the UV region. The maximum peak is located at 9.6 eV, with a trace spotted at 11.5 eV in the UV region. These intense absorption peaks arise due to light-matter interactions, inducing shifts in the valence and conduction bands. The smaller absorption peaks correspond to electron transfer in the UV region. The majority of the high-intensity absorption peaks are positioned in the UV region, suggesting that CaV$_2$TeO$_8$ is a suitable candidate for UV-operating optoelectronic devices.

The physical representation of electron movement and matter collision is described by the electron loss function, $L(\omega)$. The electron loss function is calculated \cite{Linear2006} from the following relation \ref{eq:22}:
\begin{equation}
	L(\omega) = \frac{\varepsilon_i(\omega)}{\varepsilon_r^2(\omega) + \varepsilon_i^2(\omega)} \tag{22} \quad \label{eq:22}
\end{equation}
The energy loss spectra against photon energy are plotted in Fig\ref{fig:optical_properties}(f). A first minor peak is observed at 4.2 eV with an energy loss of 0.2. The second intense peak is positioned at 10.30 eV, and the third maximum intense peak is located at 12.20 eV, with energy losses of 0.55 and 0.7, respectively. The results of the energy loss and absorption curves indicate minimum energy loss over a wide absorption range. Further, the refractive index $\eta(\omega)$, describing the transmission of incident light through CaV$_2$TeO$_8$, is calculated \cite{Linear2006} using the following formula \ref{eq:23}:
\begin{equation}
	\eta(\omega) = \frac{\sqrt{\varepsilon_r^2(\omega) + \varepsilon_i^2(\omega) - \varepsilon_r(\omega)}}{\sqrt{2}} \tag{23} \quad \label{eq:23}
\end{equation}
The refractive index is inversely proportional to the incident light absorption rate. The static refractive index $\eta_0(\omega)$ is observed at 0 eV with a value of 1.65. Additionally, two sharp peaks in the refractive index are noted at 4.1 eV and 7.2 eV, with values of 2.25 and 2.23, respectively. A gradual decrement in refractive index values is observed after 7.4 eV.
\subsection{Transport properties}
In order to understand the mobility of photoinduced carriers of exciton were investigated by utilizing the Bardeen and Shockley’s deformation potential theory ~\cite{Bardeen1950,Gao2021} of carrier mobility \ref{eq:1}. The effective mass of charge carriers plays an essential role in determining carrier mobility. Generally, the effective mass of an electron is higher than that of a hole. The trend in the effective mass of electrons directly correlates with valence band dispersion rather than the conduction band. The effective masses for various band transitions were calculated and summarized in Table~\ref{tab:exciton_properties}.

In addition to the effective mass, the elastic modulus ($C_\mathrm{3D}$) and deformation potential ($E_{ij}$) corroborate carrier mobility. The $E_{ij}$ values were calculated from the linear implementation of lattice strain ($-0.06$ to $0.06$) with the VBM and CBM band edges, as shown in Fig.~\ref{fig:cbm_vbm_analysis}. The calculated $E_{ij}$ and $C_\mathrm{3D}$ values are summarized in Table \ref{tab:carrier_mobility}. 
For the direct transition between T$–$T, the effective masses of electrons and holes were observed to be 1.74 and 1.73, respectively. Similarly, for the indirect band transition Y-$\Gamma$, the electron and hole effective masses were 0.65 and 0.94, respectively. The smaller values of effective masses for the indirect band transition indicate higher carrier mobilities of $2409.91~\mathrm{cm^2 V^{-1} s^{-1}}$ for electrons and $316.46~\mathrm{cm^2 V^{-1} s^{-1}}$ for holes.

\begin{figure}[htbp]
	\centering
	\includegraphics[width=0.35\textwidth]{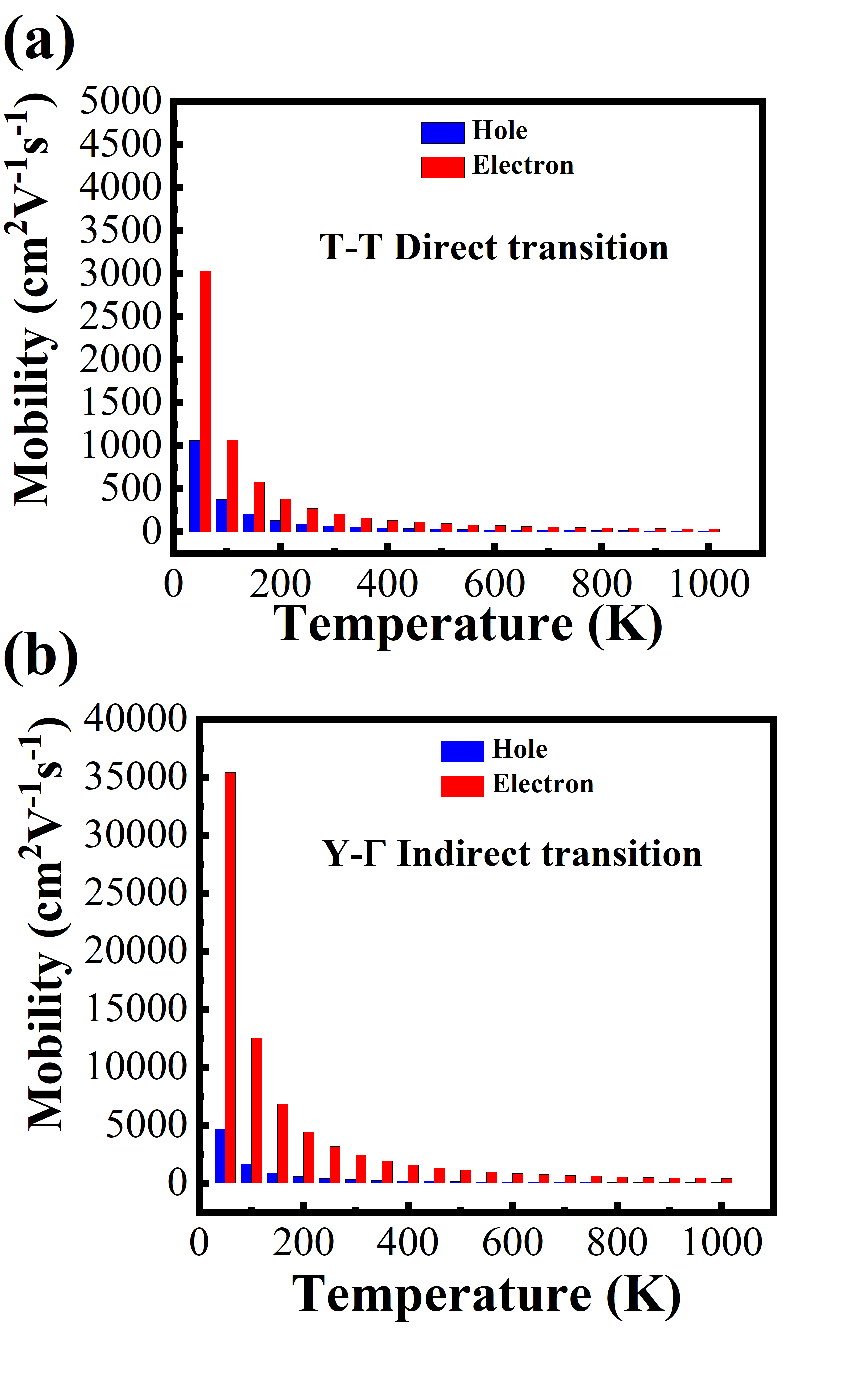}
	\vspace{-25pt}
	\caption{Carrier mobility versus temperature for (a) T-T Direct transition, and (b) Y-$\Gamma$ Indirect transition.}
	\label{fig:carrier_mobility}
\end{figure}

Conversely, the T$–$T transition showed lower carrier mobilities of $206.04~\mathrm{cm^2 V^{-1} s^{-1}}$ and $72.34~\mathrm{cm^2 V^{-1} s^{-1}}$ for electrons and holes, respectively. The carrier mobility of CaV$_2$TeO$_8$ as a function of temperature is illustrated in Fig.~\ref{fig:carrier_mobility}. The carrier mobility of CaV$_2$TeO$_8$ is comparable to that of several optoelectronic benchmark materials such as 3C-SiC, CsPbI$_3$, T-Carbon, and GaAs \cite{DGao2021, Meng2019, Sun2019,Balaghi2021}. The figure exhibits a linearly decreasing trend in carrier mobility with increasing temperature.
\begin{table}[htbp]
	\caption{\label{tab:carrier_mobility} Carrier mobility, deformation potential, and elastic modulus ($C_{3D}$) for T-T and Y-$\Gamma$ transitions in CaV$_2$TeO$_8$.}
	\begin{ruledtabular}
		\begin{tabular}{cccccc}
			Transition & \multicolumn{2}{c}{Carrier Mobility} & \multicolumn{2}{c}{Deformation Potential}\\ &&&&$E_{ij}$ & $C_{3D}$ \\
			& \multicolumn{2}{c}{(cm$^2$~(Vs)$^{-1}$)} & \multicolumn{2}{c}{($\times 10^{-18}$~J)} & (GPa) \\
			\cline{2-3} \cline{4-5}
			& $e^-$ & $h^+$ & $e^-$ & $h^+$ &  \\
			\hline
			T-T        & 206.04  & 72.34   & 6.402  & 1.069  & 215.73 \\
			Y-$\Gamma$ & 2409.91 & 316.46  & 6.350  & 1.101  & ---    \\
		\end{tabular}
	\end{ruledtabular}
\end{table}

\subsection{Thermoelectric properties}

\begin{figure}[htbp]
	\centering
	\includegraphics[width=1\linewidth]{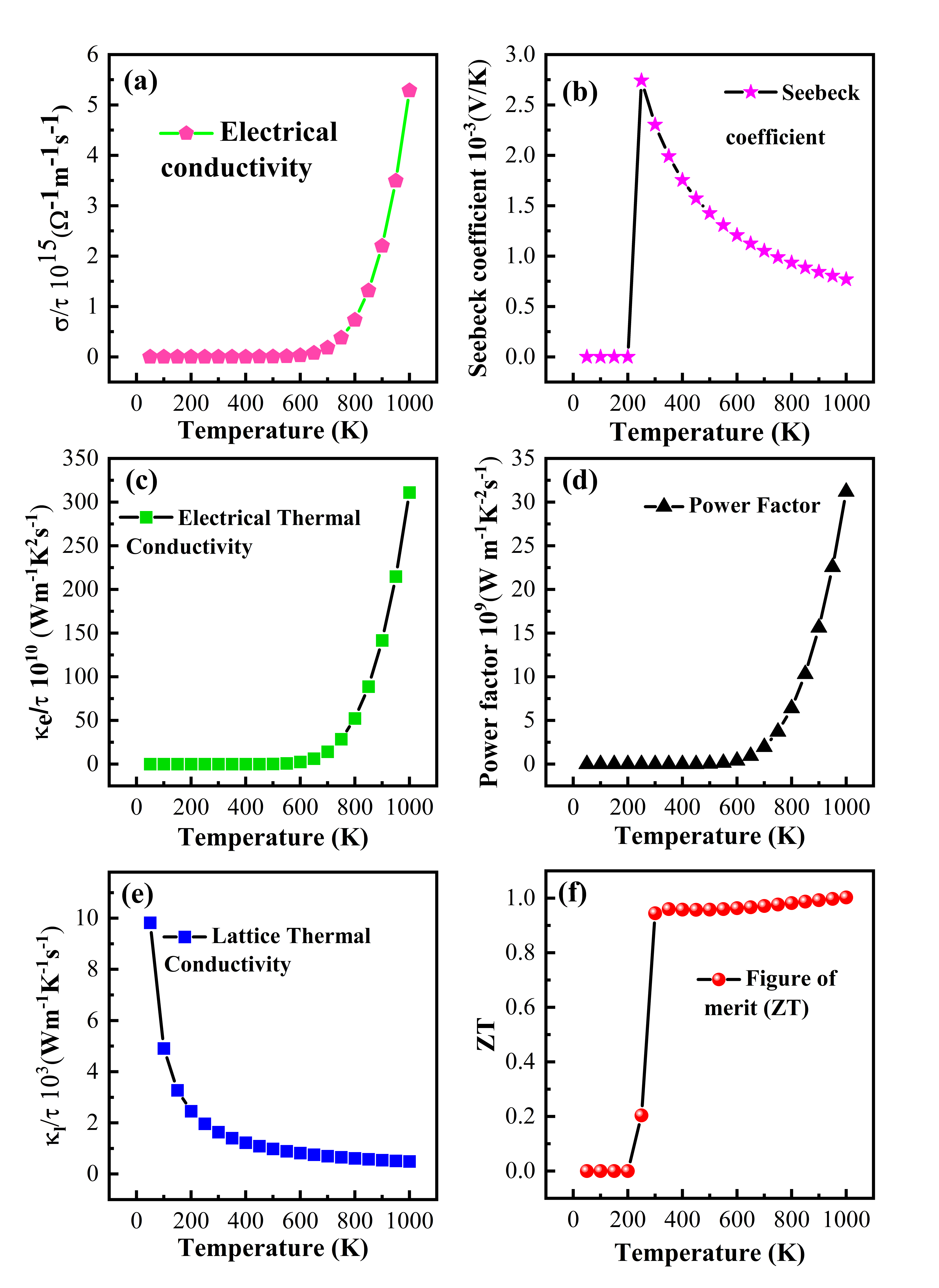}
	\caption{Temperature-dependent thermoelectric properties of CaV$_2$TeO$_8$: (a) Electrical conductivity, (b) Seebeck coefficient, (c) Electrical thermal conductivity, and (d) Power factor.}
	\label{fig:boltz}
\end{figure}

The thermoelectric properties of CaV$_2$TeO$_8$ were investigated using Boltzmann transport theory with the BoltzTraP code~\cite{MADSEN2006}. As thermoelectric energy conversion is concern, the seebeck coefficient (voltage response to temperature gradient), electrical conductance, electronic heat conductance (as an influence of relaxation time $\tau = 10^{-14}$ s) and power factor are indispensable components. The thermoelectric properties were investigated within the temperature range of 50 to 1000~K, as illustrated in Fig.~\ref{fig:boltz}. The Fig.~\ref{fig:boltz}(a) shows the electrical conductivity ($\sigma/\tau$) as a function of relaxation time in response to temperature. At room temperature (300~K), CaV$_2$TeO$_8$ attains an electrical conductivity of $5.06 \times 10^{18}$~(1/$\Omega$ms). As the temperature increases, the electrical conductivity of CaV$_2$TeO$_8$ shows an increasing trend, as seen in Fig.~\ref{fig:boltz}(b). The maximum electrical conductivity of CaV$_2$TeO$_8$ is $5.2 \times 10^{15}$~(1/$\Omega$ms) at 1000~K.
To understand the thermoelectric voltage, the Seebeck coefficient is shown in Fig. ~\ref{fig:boltz}(c). It starts with a high value of 2743~$\mu$V/K at 250~K and decreases to 2304 $\mu$V/K at 300 K. A gradual decrement in the Seebeck coefficient as a function of temperature is observed, with the maximum value of 2743 $\mu$V/K at 250 K indicating a temperature dependent gradient. 
Fig.~\ref{fig:boltz}(d) shows the electrical thermal conductivity per relaxation time ($\kappa_e/\tau$) as a function of temperature. Thermal conductivity is defined as the heat transfer due to thermally generated carriers. A maximum value of $3.11 \times 10^{12}$ (W/mKs) is obtained at 1000 K. The power factor is a significant parameter for understanding CaV$_2$TeO$_8$'s ability to generate electricity from external heat. The investigated power factor per relaxation time ($S^2 \sigma/\tau$) as a function of temperature is shown in Fig.~\ref{fig:boltz}(e). A high power factor of $3.12 \times 10^{10}$ (W/mK$^2$s) was observed at 1000 K.
To understand the thermoelectric power conversion efficiency, the lattice thermal conductivity ($\kappa_l$) is a crucial parameter. Lattice thermal conductivity refers to heat conduction through lattice vibration in the material. Lattice thermal conductivity was calculated using Slack's~\cite{Slack1973} equation~\ref{eq:24}:
\begin{equation}
	\kappa_l = A \frac{M_a \theta_a^3 \delta}{\gamma^2 T n^{2/3}} \tag{24} \quad \label{eq:24}
\end{equation}
Here $\theta_a$ refers acoustic Debye temperature, while $\delta$ corresponds to the cube root of the average volume per atoms in the molecule. Further, $\gamma$ states the Grüneisen parameter, $M$ is the average mass per atom, $n$ is the number of atoms per cell, and $A$ refers physical quantity. The physical quantity $A$ is given by:
\begin{equation}
	A = \frac{2.43 \times 10^{-8}}{1 - 0.514/\gamma + 0.228/\gamma^2} \tag{25}
\end{equation}
The acoustic Debye temperature is obtained from the relation $\theta_a = \theta_e / \sqrt[3]{n}$, where $\theta_e$ is the elastic Debye temperature. The transverse and longitudinal wave velocities, $v_t$ and $v_l$, were calculated from the IRelast code interfaced with WIEN2k~\cite{Jamal2018}. The Grüneisen parameter was calculated using~\ref{eq:26}:
\begin{equation}
	\gamma = \frac{9 - 12 (v_t/v_l)^2}{2 + 4 (v_t/v_l)^2} \tag{26} \quad \label{eq:26}
\end{equation}
The calculated elastic and acoustic Debye temperatures are $\theta_e$ = 632.5~K and $\theta_a$ = 398.4~K, respectively. The calculated Grüneisen parameter is 1.63. At 300~K and 1000~K, the lattice thermal conductivity per relaxation time ($\kappa_l/\tau$) is 1637.2 and 491.1~(W/mK$^2$s), respectively. A high lattice thermal conductivity per relaxation time of 9823.63~(W/mKs) was observed at 50~K.The figure of merit ($ZT$) is a dimensionless parameter that determines the performance of thermoelectric devices~\cite{Ahmed2024}. The thermoelectric performance of CaV$_2$TeO$_8$ was calculated from the following relation \ref{eq:27}:
\begin{equation}
	ZT = \frac{S^2 \sigma T}{\kappa_l + \kappa_e} \tag{27} \quad \label{eq:27}
\end{equation}
The $ZT$ of CaV$_2$TeO$_8$ was found to be 0.94 at 300 K and 1 at 1000~K. The significant value of $ZT$ 0.94 at room temperature  shows that CaV$_2$TeO$_8$ is a suitable candidate for flexible thermoelectric device fabrication.
\\
\section{Conclusion}
a first principles Full-potential agumented plane wave implemented DFT were utilized to calculate the structural,  mechanical and elastic as well optical, electronic and transport, thermometric properties of novel heteroanionic CaV$_2$TeO$_8$ . The structural properties were calculated using the Birch-Murnaghan equation of state and mechanical stability analysis. The calculated lattice parameters were correlated with reported experimental structural insights. Mechanical investigation of nine elastic  constants of CaV$_2$TeO$_8$ satisfies the Born criteria of mechanical and structural stability.  a Zener's anisotropic factor of 1.19 with stretching deformation in its structure. Further, the anisotropic ratio of  pugh's and poissons, with Cauchy pressure analysis uncovers the nature of ductile. Electronic bandstructure investigation, and density of states of CaV$_2$TeO$_8$ exhibits both direct and indirect band transitions with a wide band gap of 2.8~eV, showing semiconducting behavior with PBE-GGA exchange correlation. Excited electron and hole binding energies in the Y-$\Gamma$ and T-T transitions were found to be Frenkel-type strong exciton. In precises, charge carrier mobility of the Y-$\Gamma$ transition were found to be high. Optically, CaV$_2$TeO$_8$ demonstrates good absorption properties in the visible and UV ranges (2.8 to 3.8~eV), making it a moderate candidate for photocatalysis in solar water splitting. To support this, band edge calculations were carried out with reference to the NHE from pH 0 to 12. CaV$_2$TeO$_8$ exhibits photocatalytic activity within acidic media. Furthermore, the significant values of carrier mobility and wide band gap make CaV$_2$TeO$_8$ a promising photocatalyst for hydrogen evolution reactions. Additionally, the high Seebeck coefficient, ZT, and power factor at room temperature suggest that CaV$_2$TeO$_8$ is a suitable candidate for thermoelectric applications. In midst of thermometric parameters a temperature gradiance of mobility and debye temperature were investigated for the modest ZT values in room temperature(300 K) with $~$0.94. Predicted ZT, Thermal conductivity, Power factor and mobility of  CaV$_2$TeO$_8$ auspicious to waste heat recovery thermoelectric application. 
\appendix
\section{Supplementary Figures and Table}
\begin{table}[htbp]
	\caption{Fermi level and band gap as a function of strain.}
	\begin{ruledtabular}
		\begin{tabular}{r|cc}
			\text{Strain (\%)} & \text{Fermi Level (eV)} & \text{Band Gap (eV)} \\ \hline
			-0.06 & 4.03 & 2.67 \\
			-0.04 & 4.00 & 2.70 \\
			-0.02 & 3.98 & 2.71 \\
			0.00 & 3.97 & 2.80 \\
			0.02 & 3.96 & 2.73 \\
			0.04 & 3.95 & 2.67 \\
			0.06 & 3.96 & 2.65 \\
		\end{tabular}
	\end{ruledtabular}
\end{table}
\begin{table}[htbp]
	\caption{CBM and VBM edges as a function of strain for R-R and \(\text{Y-S} \mid \Gamma\) directions.}
	\begin{ruledtabular}
		\begin{tabular}{rcccc}
			\text{Strain (\%)} & \multicolumn{2}{c}{\text{R-R}} & \multicolumn{2}{c}{\(\text{Y-S} \mid \Gamma\)} \\
			& \text{CBM} & \text{VBM} & \text{CBM} & \text{VBM} \\
			\hline
			-0.06 & 2.680 & 0.009 & 2.835 & -0.010 \\
			-0.04 & 2.709 & 0.002 & 2.916 &  0.016 \\
			-0.02 & 2.730 & 0.002 & 2.960 &  0.010 \\
			0.00 & 2.720 & 0.011 & 2.825 & -0.040 \\
			0.02 & 2.723 & 0.027 & 2.947 & -0.023 \\
			0.04 & 2.698 & 0.020 & 2.879 & -0.047 \\
			0.06 & 2.655 & 0.006 & 2.869 & -0.044 \\
		\end{tabular}
	\end{ruledtabular}
\end{table}

\begin{table}[htbp]
	\caption{\label{tab:RT_TT_edges} CBM and VBM edges as a function of strain for R-T and T-T directions.}
	\begin{ruledtabular}
		\begin{tabular}{rcccc}
			\text{Strain (\%)} & \multicolumn{2}{c}{\text{R-T transition}} & \multicolumn{2}{c}{\text{T-T transition}} \\
			& \text{CBM} & \text{VBM} & \text{CBM} & \text{VBM} \\ \hline
			-0.06 & 2.679 & -0.006 & 2.679 & -0.006 \\
			-0.04 & 2.740 &  0.010 & 2.716 &  0.010 \\
			-0.02 & 2.744 &  0.016 & 2.723 &  0.016 \\
			0.00 & 2.689 &  0.020 & 2.662 &  0.020 \\
			0.02 & 2.747 &  0.016 & 2.682 &  0.016 \\
			0.04 & 2.709 &  0.016 & 2.709 &  0.016 \\
			0.06 & 2.665 &  0.017 & 2.665 &  0.017 \\
		\end{tabular}
	\end{ruledtabular}
\end{table}
\begin{figure}[htbp]
	\centering
	\includegraphics[width=0.5\textwidth]{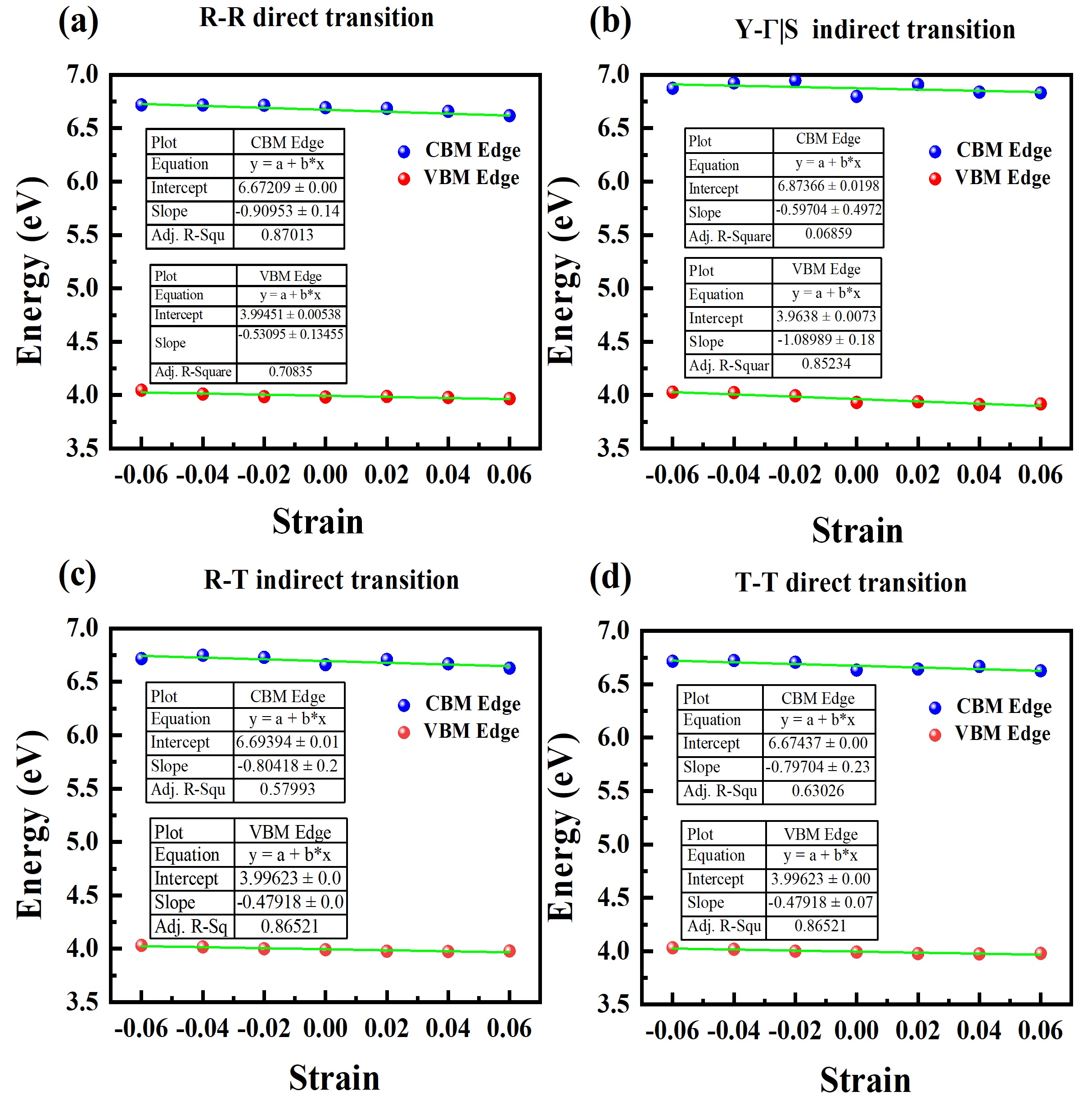} 
\caption{Deformation potential analysis of the CBM and VBM energy transitions for different strain values. (a) R-R direct transition, (b) Y-Γ indirect transition, (c) R-T indirect transition, and (d) T-T direct transition. The equations, intercepts, slopes, and adjusted R-squared values are annotated for each case.}
	\label{fig:cbm_vbm_analysis}
\end{figure}
\begin{figure}[htbp]
	\centering
	\includegraphics[width=0.49\textwidth]{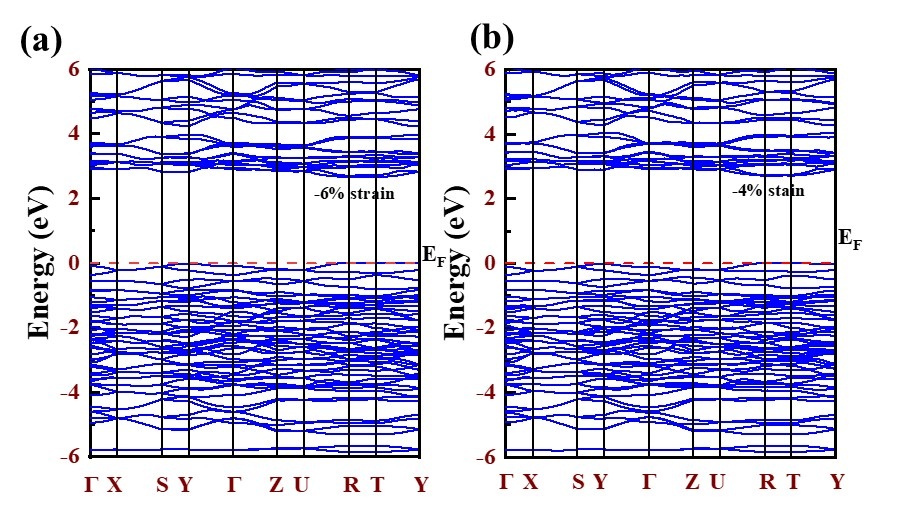}
	\includegraphics[width=0.49\textwidth]{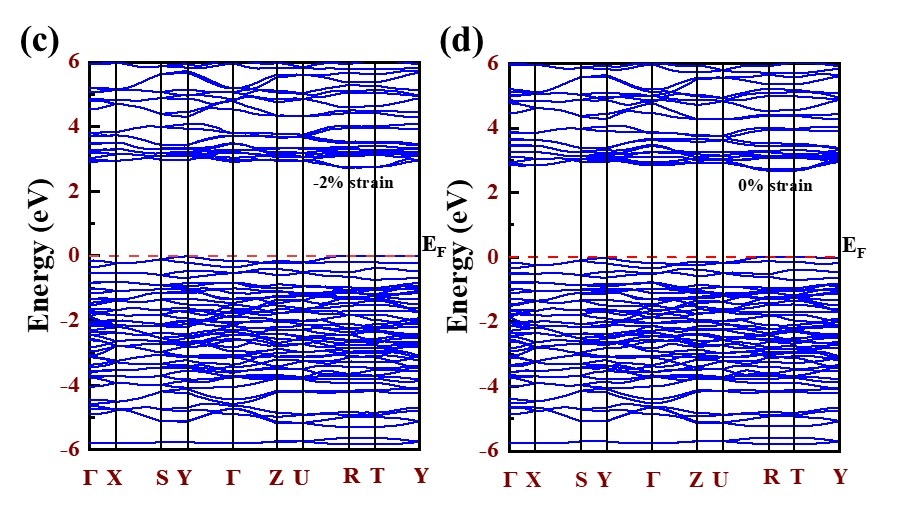}
	\includegraphics[width=0.49\textwidth]{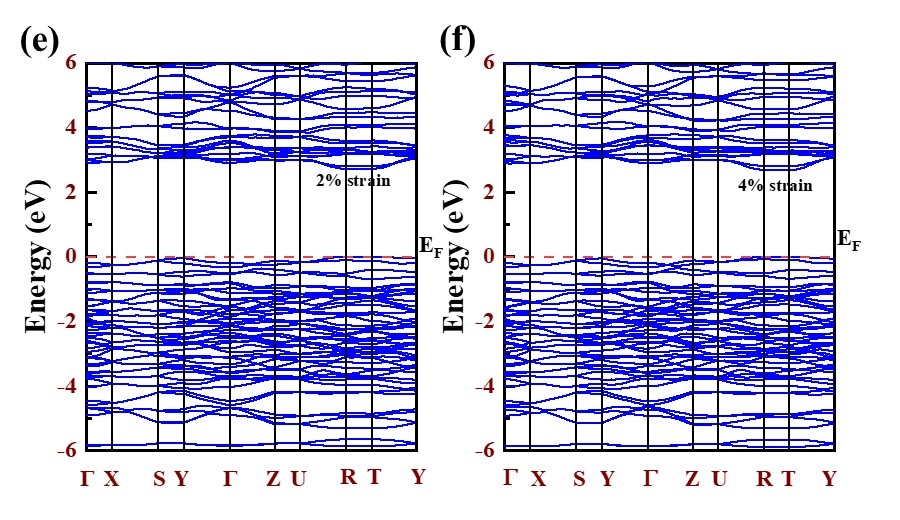}
	\includegraphics[width=0.49\textwidth]{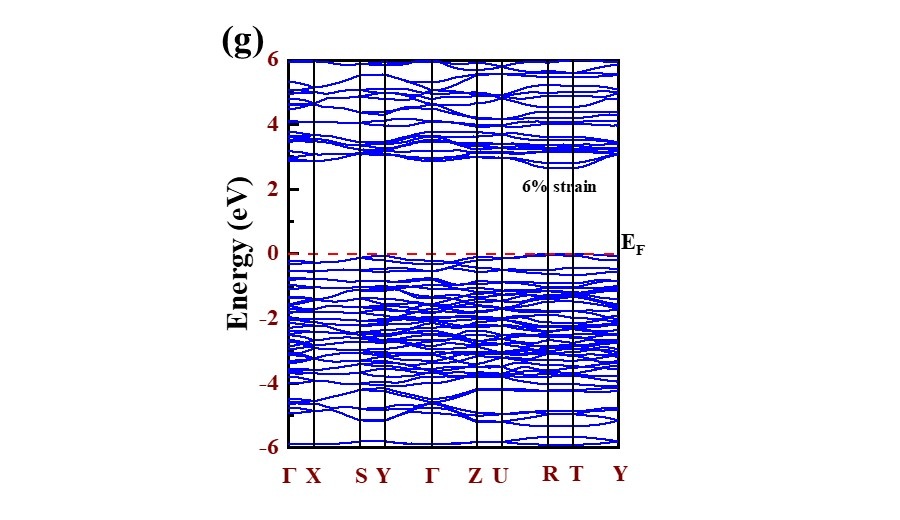} 
	\caption{Electronic band structure for bi-axial strain conditions: (a) -6\% strain (b) -4\% strain (c) -2 \% Strain (d) 0\% strain (e) +2\% strain (f) +4 \% strain and (g) +6 \% strain energy levels at first Brillouin zone points.}
	\label{fig:strain_induced_band_structure} 
\end{figure}
\begin{figure}[htbp]
		\centering
		\includegraphics[width=0.45\textwidth]{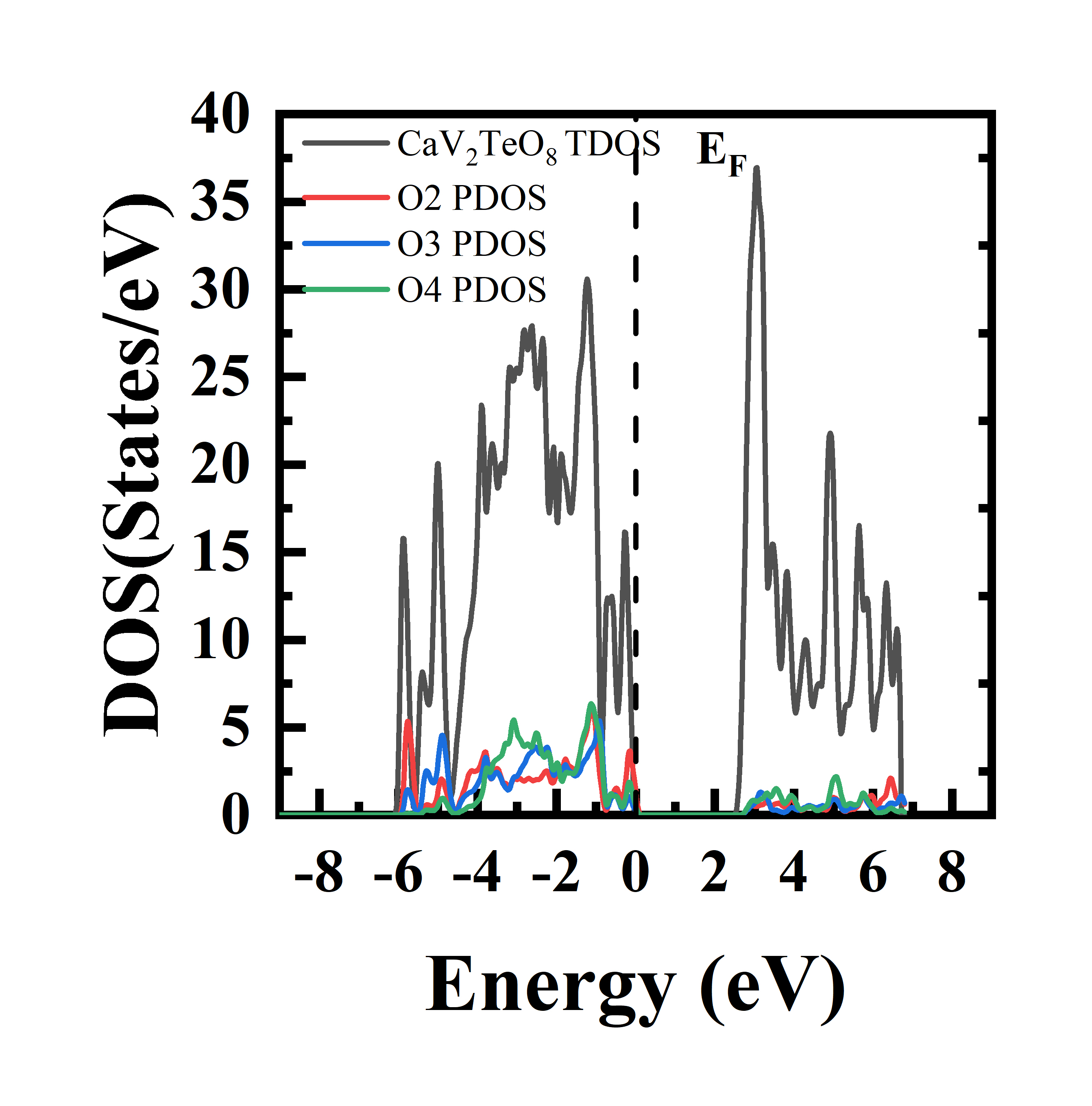} 
		\caption{Density of States (DOS) for Oxygen atoms.}
		\label{fig:O_DOS}
	\end{figure}
\clearpage
\newpage
\bibliographystyle{apsrev4-2} 
\bibliography{manuscript} 

\begin{thebibliography}{44}%
\makeatletter
\providecommand \@ifxundefined [1]{%
 \@ifx{#1\undefined}
}%
\providecommand \@ifnum [1]{%
 \ifnum #1\expandafter \@firstoftwo
 \else \expandafter \@secondoftwo
 \fi
}%
\providecommand \@ifx [1]{%
 \ifx #1\expandafter \@firstoftwo
 \else \expandafter \@secondoftwo
 \fi
}%
\providecommand \natexlab [1]{#1}%
\providecommand \enquote  [1]{``#1''}%
\providecommand \bibnamefont  [1]{#1}%
\providecommand \bibfnamefont [1]{#1}%
\providecommand \citenamefont [1]{#1}%
\providecommand \href@noop [0]{\@secondoftwo}%
\providecommand \href [0]{\begingroup \@sanitize@url \@href}%
\providecommand \@href[1]{\@@startlink{#1}\@@href}%
\providecommand \@@href[1]{\endgroup#1\@@endlink}%
\providecommand \@sanitize@url [0]{\catcode `\\12\catcode `\$12\catcode
  `\&12\catcode `\#12\catcode `\^12\catcode `\_12\catcode `\%12\relax}%
\providecommand \@@startlink[1]{}%
\providecommand \@@endlink[0]{}%
\providecommand \url  [0]{\begingroup\@sanitize@url \@url }%
\providecommand \@url [1]{\endgroup\@href {#1}{\urlprefix }}%
\providecommand \urlprefix  [0]{URL }%
\providecommand \Eprint [0]{\href }%
\providecommand \doibase [0]{https://doi.org/}%
\providecommand \selectlanguage [0]{\@gobble}%
\providecommand \bibinfo  [0]{\@secondoftwo}%
\providecommand \bibfield  [0]{\@secondoftwo}%
\providecommand \translation [1]{[#1]}%
\providecommand \BibitemOpen [0]{}%
\providecommand \bibitemStop [0]{}%
\providecommand \bibitemNoStop [0]{.\EOS\space}%
\providecommand \EOS [0]{\spacefactor3000\relax}%
\providecommand \BibitemShut  [1]{\csname bibitem#1\endcsname}%
\let\auto@bib@innerbib\@empty
\bibitem [{\citenamefont {Fu}\ \emph {et~al.}(2024)\citenamefont {Fu},
  \citenamefont {Cao}, \citenamefont {Zhang}, \citenamefont {Yang},
  \citenamefont {Zhu}, \citenamefont {Liang}, \citenamefont {Li}, \citenamefont
  {Sun},\ and\ \citenamefont {Li}}]{Fu2024}%
  \BibitemOpen
  \bibfield  {author} {\bibinfo {author} {\bibfnamefont {R.}~\bibnamefont
  {Fu}}, \bibinfo {author} {\bibfnamefont {X.}~\bibnamefont {Cao}}, \bibinfo
  {author} {\bibfnamefont {H.}~\bibnamefont {Zhang}}, \bibinfo {author}
  {\bibfnamefont {L.}~\bibnamefont {Yang}}, \bibinfo {author} {\bibfnamefont
  {Z.}~\bibnamefont {Zhu}}, \bibinfo {author} {\bibfnamefont {W.}~\bibnamefont
  {Liang}}, \bibinfo {author} {\bibfnamefont {J.}~\bibnamefont {Li}}, \bibinfo
  {author} {\bibfnamefont {H.}~\bibnamefont {Sun}},\ and\ \bibinfo {author}
  {\bibfnamefont {A.}~\bibnamefont {Li}},\ }\href
  {https://doi.org/https://doi.org/10.1016/j.seppur.2023.125283} {\bibfield
  {journal} {\bibinfo  {journal} {Sep. Purif. Technol.}\ }\textbf {\bibinfo
  {volume} {330}},\ \bibinfo {pages} {125283} (\bibinfo {year}
  {2024})}\BibitemShut {NoStop}%
\bibitem [{\citenamefont {Kannan}\ and\ \citenamefont
  {Vakeesan}(2016)}]{Kannan2016}%
  \BibitemOpen
  \bibfield  {author} {\bibinfo {author} {\bibfnamefont {N.}~\bibnamefont
  {Kannan}}\ and\ \bibinfo {author} {\bibfnamefont {D.}~\bibnamefont
  {Vakeesan}},\ }\href
  {https://doi.org/https://doi.org/10.1016/j.rser.2016.05.022} {\bibfield
  {journal} {\bibinfo  {journal} {Renew. Sustain. Energy Rev.}\ }\textbf
  {\bibinfo {volume} {62}},\ \bibinfo {pages} {1092} (\bibinfo {year}
  {2016})}\BibitemShut {NoStop}%
\bibitem [{\citenamefont {Hanif}\ \emph {et~al.}(2019)\citenamefont {Hanif},
  \citenamefont {{Faraz Raza}}, \citenamefont {de~Santos},\ and\ \citenamefont
  {Abbas}}]{Hanif2019}%
  \BibitemOpen
  \bibfield  {author} {\bibinfo {author} {\bibfnamefont {I.}~\bibnamefont
  {Hanif}}, \bibinfo {author} {\bibfnamefont {S.~M.}\ \bibnamefont {{Faraz
  Raza}}}, \bibinfo {author} {\bibfnamefont {P.~G.}\ \bibnamefont
  {de~Santos}},\ and\ \bibinfo {author} {\bibfnamefont {Q.}~\bibnamefont
  {Abbas}},\ }\href
  {https://doi.org/https://doi.org/10.1016/j.energy.2019.01.011} {\bibfield
  {journal} {\bibinfo  {journal} {Energy}\ }\textbf {\bibinfo {volume} {171}},\
  \bibinfo {pages} {493} (\bibinfo {year} {2019})}\BibitemShut {NoStop}%
\bibitem [{\citenamefont {Fujishima}\ and\ \citenamefont
  {Honda}(1972)}]{Fujishima1972}%
  \BibitemOpen
  \bibfield  {author} {\bibinfo {author} {\bibfnamefont {A.}~\bibnamefont
  {Fujishima}}\ and\ \bibinfo {author} {\bibfnamefont {K.}~\bibnamefont
  {Honda}},\ }\href {https://doi.org/10.1038/238037a0} {\bibfield  {journal}
  {\bibinfo  {journal} {Nature}\ }\textbf {\bibinfo {volume} {238}},\ \bibinfo
  {pages} {37} (\bibinfo {year} {1972})}\BibitemShut {NoStop}%
\bibitem [{\citenamefont {Negedu}\ \emph {et~al.}(2022)\citenamefont {Negedu},
  \citenamefont {Tromer}, \citenamefont {Siddique}, \citenamefont {Woellner},
  \citenamefont {Olu}, \citenamefont {Palit}, \citenamefont {Roy},
  \citenamefont {Pandey}, \citenamefont {Galvao}, \citenamefont {Kumbhakar},\
  and\ \citenamefont {Tiwary}}]{Negedu2022}%
  \BibitemOpen
  \bibfield  {author} {\bibinfo {author} {\bibfnamefont {S.~D.}\ \bibnamefont
  {Negedu}}, \bibinfo {author} {\bibfnamefont {R.}~\bibnamefont {Tromer}},
  \bibinfo {author} {\bibfnamefont {S.}~\bibnamefont {Siddique}}, \bibinfo
  {author} {\bibfnamefont {C.~F.}\ \bibnamefont {Woellner}}, \bibinfo {author}
  {\bibfnamefont {F.~E.}\ \bibnamefont {Olu}}, \bibinfo {author} {\bibfnamefont
  {M.}~\bibnamefont {Palit}}, \bibinfo {author} {\bibfnamefont {A.~K.}\
  \bibnamefont {Roy}}, \bibinfo {author} {\bibfnamefont {P.}~\bibnamefont
  {Pandey}}, \bibinfo {author} {\bibfnamefont {D.~S.}\ \bibnamefont {Galvao}},
  \bibinfo {author} {\bibfnamefont {P.}~\bibnamefont {Kumbhakar}},\ and\
  \bibinfo {author} {\bibfnamefont {C.~S.}\ \bibnamefont {Tiwary}},\ }\href
  {https://doi.org/10.1007/s00339-022-05425-z} {\bibfield  {journal} {\bibinfo
  {journal} {Appl. Phys. A}\ }\textbf {\bibinfo {volume} {128}},\ \bibinfo
  {pages} {379} (\bibinfo {year} {2022})}\BibitemShut {NoStop}%
\bibitem [{\citenamefont {Lu}\ \emph {et~al.}(2023)\citenamefont {Lu},
  \citenamefont {Luo}, \citenamefont {Dong}, \citenamefont {Ge}, \citenamefont
  {Han}, \citenamefont {Liu}, \citenamefont {Xue}, \citenamefont {Ma},
  \citenamefont {Huang}, \citenamefont {Zhou},\ and\ \citenamefont
  {Xu}}]{Lu2023}%
  \BibitemOpen
  \bibfield  {author} {\bibinfo {author} {\bibfnamefont {C.}~\bibnamefont
  {Lu}}, \bibinfo {author} {\bibfnamefont {M.}~\bibnamefont {Luo}}, \bibinfo
  {author} {\bibfnamefont {W.}~\bibnamefont {Dong}}, \bibinfo {author}
  {\bibfnamefont {Y.}~\bibnamefont {Ge}}, \bibinfo {author} {\bibfnamefont
  {T.}~\bibnamefont {Han}}, \bibinfo {author} {\bibfnamefont {Y.}~\bibnamefont
  {Liu}}, \bibinfo {author} {\bibfnamefont {X.}~\bibnamefont {Xue}}, \bibinfo
  {author} {\bibfnamefont {N.}~\bibnamefont {Ma}}, \bibinfo {author}
  {\bibfnamefont {Y.}~\bibnamefont {Huang}}, \bibinfo {author} {\bibfnamefont
  {Y.}~\bibnamefont {Zhou}},\ and\ \bibinfo {author} {\bibfnamefont
  {X.}~\bibnamefont {Xu}},\ }\href {https://doi.org/10.1002/advs.202205460}
  {\bibfield  {journal} {\bibinfo  {journal} {Adv. Sci.}\ }\textbf {\bibinfo
  {volume} {10}},\ \bibinfo {pages} {2205460} (\bibinfo {year}
  {2023})}\BibitemShut {NoStop}%
\bibitem [{\citenamefont {Puthirath~Balan}\ \emph {et~al.}(2018)\citenamefont
  {Puthirath~Balan}, \citenamefont {Radhakrishnan}, \citenamefont {Neupane},
  \citenamefont {Yazdi}, \citenamefont {Deng}, \citenamefont {A.~de~los Reyes},
  \citenamefont {Apte}, \citenamefont {B.~Puthirath}, \citenamefont {Rao},
  \citenamefont {Paulose}, \citenamefont {Vajtai}, \citenamefont {Chu},
  \citenamefont {Martí}, \citenamefont {Varghese}, \citenamefont {Tiwary},
  \citenamefont {Anantharaman},\ and\ \citenamefont
  {Ajayan}}]{PuthirathBalan2018}%
  \BibitemOpen
  \bibfield  {author} {\bibinfo {author} {\bibfnamefont {A.}~\bibnamefont
  {Puthirath~Balan}}, \bibinfo {author} {\bibfnamefont {S.}~\bibnamefont
  {Radhakrishnan}}, \bibinfo {author} {\bibfnamefont {R.}~\bibnamefont
  {Neupane}}, \bibinfo {author} {\bibfnamefont {S.}~\bibnamefont {Yazdi}},
  \bibinfo {author} {\bibfnamefont {L.}~\bibnamefont {Deng}}, \bibinfo {author}
  {\bibfnamefont {C.}~\bibnamefont {A.~de~los Reyes}}, \bibinfo {author}
  {\bibfnamefont {A.}~\bibnamefont {Apte}}, \bibinfo {author} {\bibfnamefont
  {A.}~\bibnamefont {B.~Puthirath}}, \bibinfo {author} {\bibfnamefont {B.~M.}\
  \bibnamefont {Rao}}, \bibinfo {author} {\bibfnamefont {M.}~\bibnamefont
  {Paulose}}, \bibinfo {author} {\bibfnamefont {R.}~\bibnamefont {Vajtai}},
  \bibinfo {author} {\bibfnamefont {C.-W.}\ \bibnamefont {Chu}}, \bibinfo
  {author} {\bibfnamefont {A.~A.}\ \bibnamefont {Martí}}, \bibinfo {author}
  {\bibfnamefont {O.~K.}\ \bibnamefont {Varghese}}, \bibinfo {author}
  {\bibfnamefont {C.~S.}\ \bibnamefont {Tiwary}}, \bibinfo {author}
  {\bibfnamefont {M.~R.}\ \bibnamefont {Anantharaman}},\ and\ \bibinfo {author}
  {\bibfnamefont {P.~M.}\ \bibnamefont {Ajayan}},\ }\href
  {https://doi.org/10.1021/acsanm.8b01642} {\bibfield  {journal} {\bibinfo
  {journal} {ACS Appl. Nano Mater.}\ }\textbf {\bibinfo {volume} {1}},\
  \bibinfo {pages} {6427} (\bibinfo {year} {2018})}\BibitemShut {NoStop}%
\bibitem [{\citenamefont {Bhat}\ \emph {et~al.}(2017)\citenamefont {Bhat},
  \citenamefont {Barshilia},\ and\ \citenamefont {Nagaraja}}]{Bhat2017}%
  \BibitemOpen
  \bibfield  {author} {\bibinfo {author} {\bibfnamefont {K.~S.}\ \bibnamefont
  {Bhat}}, \bibinfo {author} {\bibfnamefont {H.~C.}\ \bibnamefont
  {Barshilia}},\ and\ \bibinfo {author} {\bibfnamefont {H.~S.}\ \bibnamefont
  {Nagaraja}},\ }\href
  {https://doi.org/https://doi.org/10.1016/j.ijhydene.2017.08.098} {\bibfield
  {journal} {\bibinfo  {journal} {Int. J. Hydrogen Energy}\ }\textbf {\bibinfo
  {volume} {42}},\ \bibinfo {pages} {24645} (\bibinfo {year}
  {2017})}\BibitemShut {NoStop}%
\bibitem [{\citenamefont {Gao}\ \emph {et~al.}(2021)\citenamefont {Gao},
  \citenamefont {Zhong}, \citenamefont {Liu}, \citenamefont {Yu},\ and\
  \citenamefont {Fan}}]{DGao2021}%
  \BibitemOpen
  \bibfield  {author} {\bibinfo {author} {\bibfnamefont {D.}~\bibnamefont
  {Gao}}, \bibinfo {author} {\bibfnamefont {W.}~\bibnamefont {Zhong}}, \bibinfo
  {author} {\bibfnamefont {Y.}~\bibnamefont {Liu}}, \bibinfo {author}
  {\bibfnamefont {H.}~\bibnamefont {Yu}},\ and\ \bibinfo {author}
  {\bibfnamefont {J.}~\bibnamefont {Fan}},\ }\href
  {https://doi.org/https://doi.org/10.1016/j.apcatb.2021.120057} {\bibfield
  {journal} {\bibinfo  {journal} {Appl. Catal. B Environ.}\ }\textbf {\bibinfo
  {volume} {290}},\ \bibinfo {pages} {120057} (\bibinfo {year}
  {2021})}\BibitemShut {NoStop}%
\bibitem [{\citenamefont {Zhang}\ \emph {et~al.}(2020)\citenamefont {Zhang},
  \citenamefont {Jia}, \citenamefont {Fan}, \citenamefont {Hu}, \citenamefont
  {Liu},\ and\ \citenamefont {Yang}}]{Zhang2020}%
  \BibitemOpen
  \bibfield  {author} {\bibinfo {author} {\bibfnamefont {Q.}~\bibnamefont
  {Zhang}}, \bibinfo {author} {\bibfnamefont {J.}~\bibnamefont {Jia}}, \bibinfo
  {author} {\bibfnamefont {J.}~\bibnamefont {Fan}}, \bibinfo {author}
  {\bibfnamefont {X.}~\bibnamefont {Hu}}, \bibinfo {author} {\bibfnamefont
  {E.}~\bibnamefont {Liu}},\ and\ \bibinfo {author} {\bibfnamefont
  {D.}~\bibnamefont {Yang}},\ }\href
  {https://doi.org/10.1016/j.ijhydene.2020.03.030} {\bibfield  {journal}
  {\bibinfo  {journal} {Int. J. Hydrogen Energy}\ }\textbf {\bibinfo {volume}
  {45}},\ \bibinfo {pages} {13340} (\bibinfo {year} {2020})}\BibitemShut
  {NoStop}%
\bibitem [{\citenamefont {Tripathi}\ and\ \citenamefont
  {Karppinen}(2021)}]{Tripathi2021}%
  \BibitemOpen
  \bibfield  {author} {\bibinfo {author} {\bibfnamefont {T.~S.}\ \bibnamefont
  {Tripathi}}\ and\ \bibinfo {author} {\bibfnamefont {M.}~\bibnamefont
  {Karppinen}},\ }\href
  {https://doi.org/https://doi.org/10.1002/admi.202100146} {\bibfield
  {journal} {\bibinfo  {journal} {Adv. Mater. Interfaces}\ }\textbf {\bibinfo
  {volume} {8}},\ \bibinfo {pages} {2100146} (\bibinfo {year}
  {2021})}\BibitemShut {NoStop}%
\bibitem [{\citenamefont {Yin}\ \emph {et~al.}(2019)\citenamefont {Yin},
  \citenamefont {Baskaran},\ and\ \citenamefont {Tiwari}}]{Yin2019}%
  \BibitemOpen
  \bibfield  {author} {\bibinfo {author} {\bibfnamefont {Y.}~\bibnamefont
  {Yin}}, \bibinfo {author} {\bibfnamefont {K.}~\bibnamefont {Baskaran}},\ and\
  \bibinfo {author} {\bibfnamefont {A.}~\bibnamefont {Tiwari}},\ }\href
  {https://doi.org/https://doi.org/10.1002/pssa.201800904} {\bibfield
  {journal} {\bibinfo  {journal} {Physica Status Solidi (A)}\ }\textbf
  {\bibinfo {volume} {216}},\ \bibinfo {pages} {1800904} (\bibinfo {year}
  {2019})}\BibitemShut {NoStop}%
\bibitem [{\citenamefont {Ebling}\ \emph {et~al.}(2007)\citenamefont {Ebling},
  \citenamefont {Jacquot}, \citenamefont {Jägle}, \citenamefont {Böttner},
  \citenamefont {Kühn},\ and\ \citenamefont {Kirste}}]{Ebling2007}%
  \BibitemOpen
  \bibfield  {author} {\bibinfo {author} {\bibfnamefont {D.~G.}\ \bibnamefont
  {Ebling}}, \bibinfo {author} {\bibfnamefont {A.}~\bibnamefont {Jacquot}},
  \bibinfo {author} {\bibfnamefont {M.}~\bibnamefont {Jägle}}, \bibinfo
  {author} {\bibfnamefont {H.}~\bibnamefont {Böttner}}, \bibinfo {author}
  {\bibfnamefont {U.}~\bibnamefont {Kühn}},\ and\ \bibinfo {author}
  {\bibfnamefont {L.}~\bibnamefont {Kirste}},\ }\href
  {https://doi.org/https://doi.org/10.1002/pssr.200701174} {\bibfield
  {journal} {\bibinfo  {journal} {Phys. Status Solidi RRL}\ }\textbf {\bibinfo
  {volume} {1}},\ \bibinfo {pages} {238} (\bibinfo {year} {2007})}\BibitemShut
  {NoStop}%
\bibitem [{\citenamefont {Schira}\ and\ \citenamefont
  {Latouche}(2020)}]{Schira2020}%
  \BibitemOpen
  \bibfield  {author} {\bibinfo {author} {\bibfnamefont {R.}~\bibnamefont
  {Schira}}\ and\ \bibinfo {author} {\bibfnamefont {C.}~\bibnamefont
  {Latouche}},\ }\href
  {https://pubs.rsc.org/en/content/articlelanding/2020/nj/d0nj02316g}
  {\bibfield  {journal} {\bibinfo  {journal} {New J. Chem.}\ }\textbf {\bibinfo
  {volume} {44}},\ \bibinfo {pages} {11602} (\bibinfo {year}
  {2020})}\BibitemShut {NoStop}%
\bibitem [{\citenamefont {Wang}\ \emph {et~al.}(2022)\citenamefont {Wang},
  \citenamefont {Liang}, \citenamefont {Zhang}, \citenamefont {Yang},\ and\
  \citenamefont {Huang}}]{Wang2022}%
  \BibitemOpen
  \bibfield  {author} {\bibinfo {author} {\bibfnamefont {R.}~\bibnamefont
  {Wang}}, \bibinfo {author} {\bibfnamefont {F.}~\bibnamefont {Liang}},
  \bibinfo {author} {\bibfnamefont {X.}~\bibnamefont {Zhang}}, \bibinfo
  {author} {\bibfnamefont {Y.}~\bibnamefont {Yang}},\ and\ \bibinfo {author}
  {\bibfnamefont {F.}~\bibnamefont {Huang}},\ }\href
  {https://pubs.rsc.org/en/content/articlelanding/2022/qi/d2qi01160c}
  {\bibfield  {journal} {\bibinfo  {journal} {Inorg. Chem. Front.}\ }\textbf
  {\bibinfo {volume} {9}},\ \bibinfo {pages} {4768} (\bibinfo {year}
  {2022})}\BibitemShut {NoStop}%
\bibitem [{\citenamefont {He}\ \emph {et~al.}(2020)\citenamefont {He},
  \citenamefont {Yao}, \citenamefont {Hegde}, \citenamefont {Naghavi},
  \citenamefont {Shen}, \citenamefont {Bushick},\ and\ \citenamefont
  {Wolverton}}]{He2020}%
  \BibitemOpen
  \bibfield  {author} {\bibinfo {author} {\bibfnamefont {J.}~\bibnamefont
  {He}}, \bibinfo {author} {\bibfnamefont {Z.}~\bibnamefont {Yao}}, \bibinfo
  {author} {\bibfnamefont {V.~I.}\ \bibnamefont {Hegde}}, \bibinfo {author}
  {\bibfnamefont {S.~S.}\ \bibnamefont {Naghavi}}, \bibinfo {author}
  {\bibfnamefont {J.}~\bibnamefont {Shen}}, \bibinfo {author} {\bibfnamefont
  {K.~M.}\ \bibnamefont {Bushick}},\ and\ \bibinfo {author} {\bibfnamefont
  {C.}~\bibnamefont {Wolverton}},\ }\href
  {https://doi.org/https://doi.org/10.1021/acs.chemmater.0c01902} {\bibfield
  {journal} {\bibinfo  {journal} {Chem. Mater.}\ }\textbf {\bibinfo {volume}
  {32}},\ \bibinfo {pages} {8229} (\bibinfo {year} {2020})}\BibitemShut
  {NoStop}%
\bibitem [{\citenamefont {Jin}\ \emph {et~al.}(2022)\citenamefont {Jin},
  \citenamefont {Ding}, \citenamefont {Qin}, \citenamefont {Li}, \citenamefont
  {Jiao}, \citenamefont {Wang}, \citenamefont {Yang},\ and\ \citenamefont
  {Lv}}]{Jin2022}%
  \BibitemOpen
  \bibfield  {author} {\bibinfo {author} {\bibfnamefont {X.}~\bibnamefont
  {Jin}}, \bibinfo {author} {\bibfnamefont {X.}~\bibnamefont {Ding}}, \bibinfo
  {author} {\bibfnamefont {Z.}~\bibnamefont {Qin}}, \bibinfo {author}
  {\bibfnamefont {Y.}~\bibnamefont {Li}}, \bibinfo {author} {\bibfnamefont
  {M.}~\bibnamefont {Jiao}}, \bibinfo {author} {\bibfnamefont {R.}~\bibnamefont
  {Wang}}, \bibinfo {author} {\bibfnamefont {X.}~\bibnamefont {Yang}},\ and\
  \bibinfo {author} {\bibfnamefont {X.}~\bibnamefont {Lv}},\ }\href
  {https://doi.org/https://doi.org/10.1021/acs.inorgchem.2c02673} {\bibfield
  {journal} {\bibinfo  {journal} {Inorg. Chem.}\ }\textbf {\bibinfo {volume}
  {61}},\ \bibinfo {pages} {17623} (\bibinfo {year} {2022})}\BibitemShut
  {NoStop}%
\bibitem [{\citenamefont {Kageyama}\ \emph {et~al.}(2018)\citenamefont
  {Kageyama}, \citenamefont {Hayashi}, \citenamefont {Maeda}, \citenamefont
  {Attfield}, \citenamefont {Hiroi}, \citenamefont {Rondinelli},\ and\
  \citenamefont {Poeppelmeier}}]{Kageyama2018}%
  \BibitemOpen
  \bibfield  {author} {\bibinfo {author} {\bibfnamefont {H.}~\bibnamefont
  {Kageyama}}, \bibinfo {author} {\bibfnamefont {K.}~\bibnamefont {Hayashi}},
  \bibinfo {author} {\bibfnamefont {K.}~\bibnamefont {Maeda}}, \bibinfo
  {author} {\bibfnamefont {J.~P.}\ \bibnamefont {Attfield}}, \bibinfo {author}
  {\bibfnamefont {Z.}~\bibnamefont {Hiroi}}, \bibinfo {author} {\bibfnamefont
  {J.~M.}\ \bibnamefont {Rondinelli}},\ and\ \bibinfo {author} {\bibfnamefont
  {K.~R.}\ \bibnamefont {Poeppelmeier}},\ }\href
  {https://doi.org/https://doi.org/10.1038/s41467-018-02838-4} {\bibfield
  {journal} {\bibinfo  {journal} {Nat. Commun.}\ }\textbf {\bibinfo {volume}
  {9}},\ \bibinfo {pages} {1} (\bibinfo {year} {2018})}\BibitemShut {NoStop}%
\bibitem [{\citenamefont {Mizuguchi}\ \emph {et~al.}(2012)\citenamefont
  {Mizuguchi}, \citenamefont {Fujihisa}, \citenamefont {Gotoh}, \citenamefont
  {Suzuki}, \citenamefont {Usui}, \citenamefont {Kuroki}, \citenamefont
  {Demura}, \citenamefont {Takano}, \citenamefont {Izawa},\ and\ \citenamefont
  {Miura}}]{Mizuguchi2013}%
  \BibitemOpen
  \bibfield  {author} {\bibinfo {author} {\bibfnamefont {Y.}~\bibnamefont
  {Mizuguchi}}, \bibinfo {author} {\bibfnamefont {H.}~\bibnamefont {Fujihisa}},
  \bibinfo {author} {\bibfnamefont {Y.}~\bibnamefont {Gotoh}}, \bibinfo
  {author} {\bibfnamefont {K.}~\bibnamefont {Suzuki}}, \bibinfo {author}
  {\bibfnamefont {H.}~\bibnamefont {Usui}}, \bibinfo {author} {\bibfnamefont
  {K.}~\bibnamefont {Kuroki}}, \bibinfo {author} {\bibfnamefont
  {S.}~\bibnamefont {Demura}}, \bibinfo {author} {\bibfnamefont
  {Y.}~\bibnamefont {Takano}}, \bibinfo {author} {\bibfnamefont
  {H.}~\bibnamefont {Izawa}},\ and\ \bibinfo {author} {\bibfnamefont
  {O.}~\bibnamefont {Miura}},\ }\href
  {https://doi.org/10.1103/PhysRevB.86.220510} {\bibfield  {journal} {\bibinfo
  {journal} {Phys. Rev. B}\ }\textbf {\bibinfo {volume} {86}},\ \bibinfo
  {pages} {220510} (\bibinfo {year} {2012})}\BibitemShut {NoStop}%
\bibitem [{\citenamefont {Pilania}\ \emph {et~al.}(2020)\citenamefont
  {Pilania}, \citenamefont {Ghosh}, \citenamefont {Hartman}, \citenamefont
  {Mishra}, \citenamefont {Stanek},\ and\ \citenamefont
  {Uberuaga}}]{Pilania2020}%
  \BibitemOpen
  \bibfield  {author} {\bibinfo {author} {\bibfnamefont {G.}~\bibnamefont
  {Pilania}}, \bibinfo {author} {\bibfnamefont {A.}~\bibnamefont {Ghosh}},
  \bibinfo {author} {\bibfnamefont {S.~T.}\ \bibnamefont {Hartman}}, \bibinfo
  {author} {\bibfnamefont {R.}~\bibnamefont {Mishra}}, \bibinfo {author}
  {\bibfnamefont {C.~R.}\ \bibnamefont {Stanek}},\ and\ \bibinfo {author}
  {\bibfnamefont {B.~P.}\ \bibnamefont {Uberuaga}},\ }\href
  {https://doi.org/https://doi.org/10.1038/s41524-020-0338-1} {\bibfield
  {journal} {\bibinfo  {journal} {Npj Comput. Mater.}\ }\textbf {\bibinfo
  {volume} {6}},\ \bibinfo {pages} {1} (\bibinfo {year} {2020})}\BibitemShut
  {NoStop}%
\bibitem [{\citenamefont {Zhao}\ \emph {et~al.}(2010)\citenamefont {Zhao},
  \citenamefont {Berardan}, \citenamefont {Pei}, \citenamefont {Byl},
  \citenamefont {Pinsard-Gaudart},\ and\ \citenamefont {Dragoe}}]{Zhao2010}%
  \BibitemOpen
  \bibfield  {author} {\bibinfo {author} {\bibfnamefont {L.~D.}\ \bibnamefont
  {Zhao}}, \bibinfo {author} {\bibfnamefont {D.}~\bibnamefont {Berardan}},
  \bibinfo {author} {\bibfnamefont {Y.~L.}\ \bibnamefont {Pei}}, \bibinfo
  {author} {\bibfnamefont {C.}~\bibnamefont {Byl}}, \bibinfo {author}
  {\bibfnamefont {L.}~\bibnamefont {Pinsard-Gaudart}},\ and\ \bibinfo {author}
  {\bibfnamefont {N.}~\bibnamefont {Dragoe}},\ }\href
  {https://doi.org/https://doi.org/10.1063/1.3485050} {\bibfield  {journal}
  {\bibinfo  {journal} {Appl. Phys. Lett.}\ }\textbf {\bibinfo {volume} {97}},\
  \bibinfo {pages} {1} (\bibinfo {year} {2010})}\BibitemShut {NoStop}%
\bibitem [{\citenamefont {Ye}\ \emph {et~al.}(2014)\citenamefont {Ye},
  \citenamefont {Su}, \citenamefont {Jin}, \citenamefont {Xie},\ and\
  \citenamefont {Zhang}}]{Ye2014}%
  \BibitemOpen
  \bibfield  {author} {\bibinfo {author} {\bibfnamefont {L.}~\bibnamefont
  {Ye}}, \bibinfo {author} {\bibfnamefont {Y.}~\bibnamefont {Su}}, \bibinfo
  {author} {\bibfnamefont {X.}~\bibnamefont {Jin}}, \bibinfo {author}
  {\bibfnamefont {H.}~\bibnamefont {Xie}},\ and\ \bibinfo {author}
  {\bibfnamefont {C.}~\bibnamefont {Zhang}},\ }\href
  {https://doi.org/https://doi.org/10.1039/C3EN00098B} {\bibfield  {journal}
  {\bibinfo  {journal} {Environ. Sci. Nano}\ }\textbf {\bibinfo {volume} {1}},\
  \bibinfo {pages} {90} (\bibinfo {year} {2014})}\BibitemShut {NoStop}%
\bibitem [{\citenamefont {Duan}\ \emph {et~al.}(2009)\citenamefont {Duan},
  \citenamefont {Delsing},\ and\ \citenamefont {Hintzen}}]{Duan2009}%
  \BibitemOpen
  \bibfield  {author} {\bibinfo {author} {\bibfnamefont {C.~J.}\ \bibnamefont
  {Duan}}, \bibinfo {author} {\bibfnamefont {A.~C.~A.}\ \bibnamefont
  {Delsing}},\ and\ \bibinfo {author} {\bibfnamefont {H.~T.}\ \bibnamefont
  {Hintzen}},\ }\href {https://pubs.acs.org/doi/abs/10.1021/cm801990r}
  {\bibfield  {journal} {\bibinfo  {journal} {ChemInform}\ }\textbf {\bibinfo
  {volume} {40}},\ \bibinfo {pages} {1010} (\bibinfo {year}
  {2009})}\BibitemShut {NoStop}%
\bibitem [{\citenamefont {Konatham}\ \emph {et~al.}(2023)\citenamefont
  {Konatham}, \citenamefont {Robert}, \citenamefont {Jaschin}, \citenamefont
  {Varma},\ and\ \citenamefont {Vidyasagar}}]{Konatham2023}%
  \BibitemOpen
  \bibfield  {author} {\bibinfo {author} {\bibfnamefont {S.}~\bibnamefont
  {Konatham}}, \bibinfo {author} {\bibfnamefont {R.}~\bibnamefont {Robert}},
  \bibinfo {author} {\bibfnamefont {P.~W.}\ \bibnamefont {Jaschin}}, \bibinfo
  {author} {\bibfnamefont {K.~B.~R.}\ \bibnamefont {Varma}},\ and\ \bibinfo
  {author} {\bibfnamefont {K.}~\bibnamefont {Vidyasagar}},\ }\href
  {https://doi.org/https://doi.org/10.1021/acs.inorgchem.3c02494} {\bibfield
  {journal} {\bibinfo  {journal} {Inorg. Chem.}\ }\textbf {\bibinfo {volume}
  {62}},\ \bibinfo {pages} {16890} (\bibinfo {year} {2023})}\BibitemShut
  {NoStop}%
\bibitem [{\citenamefont {Blaha}\ \emph {et~al.}(2020)\citenamefont {Blaha},
  \citenamefont {Schwarz}, \citenamefont {Tran}, \citenamefont {Laskowski},
  \citenamefont {Madsen},\ and\ \citenamefont {Marks}}]{Blaha2020}%
  \BibitemOpen
  \bibfield  {author} {\bibinfo {author} {\bibfnamefont {P.}~\bibnamefont
  {Blaha}}, \bibinfo {author} {\bibfnamefont {K.}~\bibnamefont {Schwarz}},
  \bibinfo {author} {\bibfnamefont {F.}~\bibnamefont {Tran}}, \bibinfo {author}
  {\bibfnamefont {R.}~\bibnamefont {Laskowski}}, \bibinfo {author}
  {\bibfnamefont {G.~K.~H.}\ \bibnamefont {Madsen}},\ and\ \bibinfo {author}
  {\bibfnamefont {L.~D.}\ \bibnamefont {Marks}},\ }\href
  {https://doi.org/10.1063/1.5143061} {\bibfield  {journal} {\bibinfo
  {journal} {J. Chem. Phys.}\ }\textbf {\bibinfo {volume} {152}},\ \bibinfo
  {pages} {074101} (\bibinfo {year} {2020})}\BibitemShut {NoStop}%
\bibitem [{\citenamefont {Perdew}\ \emph {et~al.}(1996)\citenamefont {Perdew},
  \citenamefont {Burke},\ and\ \citenamefont {Ernzerhof}}]{Perdew1996}%
  \BibitemOpen
  \bibfield  {author} {\bibinfo {author} {\bibfnamefont {J.~P.}\ \bibnamefont
  {Perdew}}, \bibinfo {author} {\bibfnamefont {K.}~\bibnamefont {Burke}},\ and\
  \bibinfo {author} {\bibfnamefont {M.}~\bibnamefont {Ernzerhof}},\ }\href
  {https://doi.org/10.1103/PhysRevLett.77.3865} {\bibfield  {journal} {\bibinfo
   {journal} {Phys. Rev. Lett.}\ }\textbf {\bibinfo {volume} {77}},\ \bibinfo
  {pages} {3865} (\bibinfo {year} {1996})}\BibitemShut {NoStop}%
\bibitem [{\citenamefont {Murnaghan}(1937)}]{Murnaghan1937}%
  \BibitemOpen
  \bibfield  {author} {\bibinfo {author} {\bibfnamefont {F.~D.}\ \bibnamefont
  {Murnaghan}},\ }\href {https://doi.org/10.2307/2371405} {\bibfield  {journal}
  {\bibinfo  {journal} {Am. J. Math.}\ }\textbf {\bibinfo {volume} {59}},\
  \bibinfo {pages} {235} (\bibinfo {year} {1937})}\BibitemShut {NoStop}%
\bibitem [{\citenamefont {Momma}\ and\ \citenamefont
  {Izumi}(2011)}]{Momma2011}%
  \BibitemOpen
  \bibfield  {author} {\bibinfo {author} {\bibfnamefont {K.}~\bibnamefont
  {Momma}}\ and\ \bibinfo {author} {\bibfnamefont {F.}~\bibnamefont {Izumi}},\
  }\href {https://doi.org/10.1107/S0021889811038970} {\bibfield  {journal}
  {\bibinfo  {journal} {J. Appl. Crystallogr.}\ }\textbf {\bibinfo {volume}
  {44}},\ \bibinfo {pages} {1272} (\bibinfo {year} {2011})}\BibitemShut
  {NoStop}%
\bibitem [{\citenamefont {Madsen}\ and\ \citenamefont
  {Singh}(2006)}]{MADSEN2006}%
  \BibitemOpen
  \bibfield  {author} {\bibinfo {author} {\bibfnamefont {G.~K.}\ \bibnamefont
  {Madsen}}\ and\ \bibinfo {author} {\bibfnamefont {D.~J.}\ \bibnamefont
  {Singh}},\ }\href {https://doi.org/10.1016/j.cpc.2006.03.007} {\bibfield
  {journal} {\bibinfo  {journal} {Comput. Phys. Commun.}\ }\textbf {\bibinfo
  {volume} {175}},\ \bibinfo {pages} {67} (\bibinfo {year} {2006})}\BibitemShut
  {NoStop}%
\bibitem [{\citenamefont {Bardeen}\ and\ \citenamefont
  {Shockley}(1950)}]{Bardeen1950}%
  \BibitemOpen
  \bibfield  {author} {\bibinfo {author} {\bibfnamefont {J.}~\bibnamefont
  {Bardeen}}\ and\ \bibinfo {author} {\bibfnamefont {W.}~\bibnamefont
  {Shockley}},\ }\href {https://doi.org/https://doi.org/10.1103/PhysRev.80.72}
  {\bibfield  {journal} {\bibinfo  {journal} {Phys. Rev.}\ }\textbf {\bibinfo
  {volume} {80}},\ \bibinfo {pages} {72} (\bibinfo {year} {1950})}\BibitemShut
  {NoStop}%
\bibitem [{\citenamefont {Wan}\ \emph {et~al.}(2023)\citenamefont {Wan},
  \citenamefont {Sun}, \citenamefont {Shi}, \citenamefont {Yan},\ and\
  \citenamefont {Zhang}}]{Wan2023}%
  \BibitemOpen
  \bibfield  {author} {\bibinfo {author} {\bibfnamefont {Y.}~\bibnamefont
  {Wan}}, \bibinfo {author} {\bibfnamefont {P.}~\bibnamefont {Sun}}, \bibinfo
  {author} {\bibfnamefont {L.}~\bibnamefont {Shi}}, \bibinfo {author}
  {\bibfnamefont {X.}~\bibnamefont {Yan}},\ and\ \bibinfo {author}
  {\bibfnamefont {X.}~\bibnamefont {Zhang}},\ }\href
  {https://doi.org/https://doi.org/10.1021/acs.jpclett.3c01850} {\bibfield
  {journal} {\bibinfo  {journal} {J. Phys. Chem. Lett.}\ }\textbf {\bibinfo
  {volume} {14}},\ \bibinfo {pages} {7411} (\bibinfo {year}
  {2023})}\BibitemShut {NoStop}%
\bibitem [{\citenamefont {Gao}\ and\ \citenamefont {Tang}(2021)}]{Gao2021}%
  \BibitemOpen
  \bibfield  {author} {\bibinfo {author} {\bibfnamefont {L.~K.}\ \bibnamefont
  {Gao}}\ and\ \bibinfo {author} {\bibfnamefont {Y.~L.}\ \bibnamefont {Tang}},\
  }\href {https://doi.org/https://doi.org/10.1021/acsomega.1c00734} {\bibfield
  {journal} {\bibinfo  {journal} {ACS Omega}\ }\textbf {\bibinfo {volume}
  {6}},\ \bibinfo {pages} {11545} (\bibinfo {year} {2021})}\BibitemShut
  {NoStop}%
\bibitem [{\citenamefont {Lau}\ and\ \citenamefont {McCurdy}(1998)}]{Lau1998}%
  \BibitemOpen
  \bibfield  {author} {\bibinfo {author} {\bibfnamefont {K.}~\bibnamefont
  {Lau}}\ and\ \bibinfo {author} {\bibfnamefont {A.}~\bibnamefont {McCurdy}},\
  }\href {https://doi.org/https://doi.org/10.1103/PhysRevB.58.8980} {\bibfield
  {journal} {\bibinfo  {journal} {Phys. Rev. B - Condens. Matter Mater. Phys.}\
  }\textbf {\bibinfo {volume} {58}},\ \bibinfo {pages} {8980} (\bibinfo {year}
  {1998})}\BibitemShut {NoStop}%
\bibitem [{\citenamefont {Liu}\ \emph {et~al.}(2022)\citenamefont {Liu},
  \citenamefont {Ekuma}, \citenamefont {Li}, \citenamefont {Yang},\ and\
  \citenamefont {Li}}]{Liu2022}%
  \BibitemOpen
  \bibfield  {author} {\bibinfo {author} {\bibfnamefont {Z.~L.}\ \bibnamefont
  {Liu}}, \bibinfo {author} {\bibfnamefont {C.~E.}\ \bibnamefont {Ekuma}},
  \bibinfo {author} {\bibfnamefont {W.~Q.}\ \bibnamefont {Li}}, \bibinfo
  {author} {\bibfnamefont {J.~Q.}\ \bibnamefont {Yang}},\ and\ \bibinfo
  {author} {\bibfnamefont {X.~J.}\ \bibnamefont {Li}},\ }\href
  {https://doi.org/https://doi.org/10.1016/j.cpc.2021.108180} {\bibfield
  {journal} {\bibinfo  {journal} {Comput. Phys. Commun.}\ }\textbf {\bibinfo
  {volume} {270}},\ \bibinfo {pages} {108180} (\bibinfo {year}
  {2022})}\BibitemShut {NoStop}%
\bibitem [{\citenamefont {Zhang}\ \emph {et~al.}(2015)\citenamefont {Zhang},
  \citenamefont {Yu}, \citenamefont {Liu},\ and\ \citenamefont
  {Liu}}]{Zhang2015}%
  \BibitemOpen
  \bibfield  {author} {\bibinfo {author} {\bibfnamefont {J.}~\bibnamefont
  {Zhang}}, \bibinfo {author} {\bibfnamefont {W.}~\bibnamefont {Yu}}, \bibinfo
  {author} {\bibfnamefont {J.}~\bibnamefont {Liu}},\ and\ \bibinfo {author}
  {\bibfnamefont {B.}~\bibnamefont {Liu}},\ }\href
  {https://doi.org/https://doi.org/10.1016/j.apsusc.2015.08.084} {\bibfield
  {journal} {\bibinfo  {journal} {Appl. Surf. Sci.}\ }\textbf {\bibinfo
  {volume} {358}},\ \bibinfo {pages} {457} (\bibinfo {year}
  {2015})}\BibitemShut {NoStop}%
\bibitem [{\citenamefont {Vijay}\ \emph {et~al.}(2022)\citenamefont {Vijay},
  \citenamefont {Hariharan}, \citenamefont {Sneha},\ and\ \citenamefont
  {Eithiraj}}]{VIJAY2022}%
  \BibitemOpen
  \bibfield  {author} {\bibinfo {author} {\bibfnamefont {A.}~\bibnamefont
  {Vijay}}, \bibinfo {author} {\bibfnamefont {M.}~\bibnamefont {Hariharan}},
  \bibinfo {author} {\bibfnamefont {G.}~\bibnamefont {Sneha}},\ and\ \bibinfo
  {author} {\bibfnamefont {R.}~\bibnamefont {Eithiraj}},\ }\href
  {https://doi.org/10.1016/j.mtcomm.2022.104952} {\bibfield  {journal}
  {\bibinfo  {journal} {Mater. Today Commun.}\ }\textbf {\bibinfo {volume}
  {33}},\ \bibinfo {pages} {104952} (\bibinfo {year} {2022})}\BibitemShut
  {NoStop}%
\bibitem [{\citenamefont {Nayak}\ and\ \citenamefont
  {Thangavel}(2021)}]{Nayak2021}%
  \BibitemOpen
  \bibfield  {author} {\bibinfo {author} {\bibfnamefont {D.}~\bibnamefont
  {Nayak}}\ and\ \bibinfo {author} {\bibfnamefont {R.}~\bibnamefont
  {Thangavel}},\ }\href {https://doi.org/10.1016/j.mseb.2020.114944} {\bibfield
   {journal} {\bibinfo  {journal} {Mater. Sci. Eng. B}\ }\textbf {\bibinfo
  {volume} {264}},\ \bibinfo {pages} {114944} (\bibinfo {year}
  {2021})}\BibitemShut {NoStop}%
\bibitem [{\citenamefont {Ambrosch-Draxl}\ and\ \citenamefont
  {Sofo}(2006)}]{Linear2006}%
  \BibitemOpen
  \bibfield  {author} {\bibinfo {author} {\bibfnamefont {C.}~\bibnamefont
  {Ambrosch-Draxl}}\ and\ \bibinfo {author} {\bibfnamefont {J.~O.}\
  \bibnamefont {Sofo}},\ }\href {https://doi.org/10.1016/j.cpc.2006.03.005}
  {\bibfield  {journal} {\bibinfo  {journal} {Comput. Phys. Commun.}\ }\textbf
  {\bibinfo {volume} {175}},\ \bibinfo {pages} {1} (\bibinfo {year}
  {2006})}\BibitemShut {NoStop}%
\bibitem [{\citenamefont {Meng}\ \emph {et~al.}(2019)\citenamefont {Meng},
  \citenamefont {Ma}, \citenamefont {He},\ and\ \citenamefont {Li}}]{Meng2019}%
  \BibitemOpen
  \bibfield  {author} {\bibinfo {author} {\bibfnamefont {F.}~\bibnamefont
  {Meng}}, \bibinfo {author} {\bibfnamefont {J.}~\bibnamefont {Ma}}, \bibinfo
  {author} {\bibfnamefont {J.}~\bibnamefont {He}},\ and\ \bibinfo {author}
  {\bibfnamefont {W.}~\bibnamefont {Li}},\ }\href
  {https://doi.org/10.1103/PhysRevB.99.045201} {\bibfield  {journal} {\bibinfo
  {journal} {Phys. Rev. B}\ }\textbf {\bibinfo {volume} {99}},\ \bibinfo
  {pages} {045201} (\bibinfo {year} {2019})}\BibitemShut {NoStop}%
\bibitem [{\citenamefont {Sun}\ \emph {et~al.}(2019)\citenamefont {Sun},
  \citenamefont {Bai}, \citenamefont {Kripalani},\ and\ \citenamefont
  {Zhou}}]{Sun2019}%
  \BibitemOpen
  \bibfield  {author} {\bibinfo {author} {\bibfnamefont {P.~P.}\ \bibnamefont
  {Sun}}, \bibinfo {author} {\bibfnamefont {L.}~\bibnamefont {Bai}}, \bibinfo
  {author} {\bibfnamefont {D.~R.}\ \bibnamefont {Kripalani}},\ and\ \bibinfo
  {author} {\bibfnamefont {K.}~\bibnamefont {Zhou}},\ }\href
  {https://doi.org/10.1038/s41524-018-0146-z} {\bibfield  {journal} {\bibinfo
  {journal} {Npj Comput. Mater.}\ }\textbf {\bibinfo {volume} {5}},\ \bibinfo
  {pages} {1} (\bibinfo {year} {2019})}\BibitemShut {NoStop}%
\bibitem [{\citenamefont {Balaghi}\ \emph {et~al.}(2021)\citenamefont {Balaghi}
  \emph {et~al.}}]{Balaghi2021}%
  \BibitemOpen
  \bibfield  {author} {\bibinfo {author} {\bibfnamefont {L.}~\bibnamefont
  {Balaghi}} \emph {et~al.},\ }\href
  {https://doi.org/10.1038/s41467-021-27006-z} {\bibfield  {journal} {\bibinfo
  {journal} {Nat. Commun.}\ }\textbf {\bibinfo {volume} {12}},\ \bibinfo
  {pages} {1} (\bibinfo {year} {2021})}\BibitemShut {NoStop}%
\bibitem [{\citenamefont {Slack}(1973)}]{Slack1973}%
  \BibitemOpen
  \bibfield  {author} {\bibinfo {author} {\bibfnamefont {G.~A.}\ \bibnamefont
  {Slack}},\ }\href {https://doi.org/10.1016/0022-3697(73)90092-9} {\bibfield
  {journal} {\bibinfo  {journal} {J. Phys. Chem. Solids}\ }\textbf {\bibinfo
  {volume} {34}},\ \bibinfo {pages} {321} (\bibinfo {year} {1973})}\BibitemShut
  {NoStop}%
\bibitem [{\citenamefont {Jamal}\ \emph {et~al.}(2018)\citenamefont {Jamal},
  \citenamefont {Bilal}, \citenamefont {Ahmad},\ and\ \citenamefont
  {Jalali-Asadabadi}}]{Jamal2018}%
  \BibitemOpen
  \bibfield  {author} {\bibinfo {author} {\bibfnamefont {M.}~\bibnamefont
  {Jamal}}, \bibinfo {author} {\bibfnamefont {M.}~\bibnamefont {Bilal}},
  \bibinfo {author} {\bibfnamefont {I.}~\bibnamefont {Ahmad}},\ and\ \bibinfo
  {author} {\bibfnamefont {S.}~\bibnamefont {Jalali-Asadabadi}},\ }\href
  {https://doi.org/10.1016/j.jallcom.2017.10.139} {\bibfield  {journal}
  {\bibinfo  {journal} {J. Alloys Compd.}\ }\textbf {\bibinfo {volume} {735}},\
  \bibinfo {pages} {569} (\bibinfo {year} {2018})}\BibitemShut {NoStop}%
\bibitem [{\citenamefont {Ahmed}\ \emph {et~al.}(2024)\citenamefont {Ahmed},
  \citenamefont {Murtaza}, \citenamefont {Irfan}, \citenamefont {Ayyaz},\ and\
  \citenamefont {Albalawi}}]{Ahmed2024}%
  \BibitemOpen
  \bibfield  {author} {\bibinfo {author} {\bibfnamefont {A.}~\bibnamefont
  {Ahmed}}, \bibinfo {author} {\bibfnamefont {G.}~\bibnamefont {Murtaza}},
  \bibinfo {author} {\bibfnamefont {M.}~\bibnamefont {Irfan}}, \bibinfo
  {author} {\bibfnamefont {A.}~\bibnamefont {Ayyaz}},\ and\ \bibinfo {author}
  {\bibfnamefont {H.}~\bibnamefont {Albalawi}},\ }\href
  {https://doi.org/10.1007/s00339-023-07235-3} {\bibfield  {journal} {\bibinfo
  {journal} {Appl. Phys. A}\ }\textbf {\bibinfo {volume} {130}},\ \bibinfo
  {pages} {66} (\bibinfo {year} {2024})}\BibitemShut {NoStop}%
\end{thebibliography}%
\end{document}